\documentclass[twocolumn,english,aps,prb,twocolum,superscriptaddress,natbib,bibnotes,amsmath,amssymb,floatfix,groupedaddress,footinbib]{revtex4-2}
\usepackage[colorlinks=true,citecolor=blue,linkcolor=magenta]{hyperref}

\usepackage[markup=nocolor, authormarkupposition=left]{changes} 

\usepackage{soul}
\usepackage[utf8]{inputenc}
\usepackage[english]{babel}
\usepackage{amsmath,amsfonts,amssymb}
\usepackage[T1]{fontenc}
\usepackage{url}

\usepackage{amsmath}
\usepackage{siunitx}
\usepackage[version=4]{mhchem}
\newcommand{\dop}{{:}}

\usepackage{amsfonts}
\usepackage{amssymb}

\usepackage{epstopdf}
\usepackage{graphicx}
\graphicspath{{./Figures/}}

\newcommand{\Fref}[1]{Figure~\ref{#1}}
\newcommand{\fref}[1]{Fig.~\ref{#1}}

\usepackage{changes}

\begin{document}

\title{A photonic integrated circuit based erbium-doped amplifier}

\author{Yang Liu$^{1,\ast,\dagger}$, 
				Zheru Qiu$^{1,\ast}$,
			    Xinru Ji$^{1}$, 
			    Jijun He$^{1}$, 
			    Johann Riemensberger$^{1}$, 
			    Martin Hafermann$^{2}$,
			    Rui Ning Wang$^{1}$,
			    Junqiu Liu$^{1}$,
				Carsten Ronning$^{2}$,
				and Tobias J.  Kippenberg$^{1,\ddag}$}
\affiliation{
$^1$Institute of Physics, Swiss Federal Institute of Technology Lausanne (EPFL), CH-1015 Lausanne, Switzerland\\
$^2$Institute of Solid State Physics, Friedrich Schiller University Jena, Max-Wien-Platz 1, 07743 Jena, Germany
}

\maketitle

\noindent\textbf{
Erbium-doped fiber amplifiers have revolutionized long-haul optical communications and laser technology. 
Erbium ions could equally provide a basis for efficient optical amplification in photonic integrated circuits.
However, this approach has thus far remained impractical due to insufficient output power.
Here, we demonstrate a photonic integrated circuit based erbium amplifier reaching 145 mW output power and more than 30 dB small-signal gain -- on par with commercial fiber amplifiers and beyond state-of-the-art III-V heterogeneously integrated semiconductor amplifiers.
We achieve this by applying ion implantation to recently emerged ultralow-loss \ce{Si3N4} photonic integrated circuits with meter-scale-length waveguides.
We utilize the device to increase by 100-fold the output power of soliton microcombs, required for low-noise photonic microwave generation or as a source for wavelength-division multiplexed optical communications.
Endowing \ce{Si3N4} photonic integrated circuits with gain enables the miniaturization of a wide range of fiber-based devices such as high-pulse-energy femtosecond mode-locked lasers.
}

\pretolerance=10000
The invention of erbium-doped fiber amplifiers (EDFAs) in the 1980s\cite{Poole1986, MEARS1987} has revolutionized long-haul optical communications and profoundly impacted our information society.  
EDFAs have replaced the complex and bandwidth-limited electrical repeaters, enabling transatlantic fiber-based optical communication networks\cite{Winzer2018}. 
Erbium amplifiers have a number of unique properties highly suitable for optical communications, such as the broadband gain around 1550 nm that coincides with the lowest optical fiber propagation loss band,  a long ms-lifetime of the parity forbidden intra-4-f shell $^4{I}_{15/2}-^4{I}_{13/2}$ transition that leads to slow gain dynamics and negligible inter-channel crosstalk in multi-wavelength amplification,  high temperature stability,  and low noise figure approaching the quantum mechanical limit of 3 dB for phase insensitive amplification\cite{Simpson1999_book}. 
Today, EDFAs have underpinned the development of narrow-linewidth and mode-locked lasers that are widely deployed in applications such as coherent communications\cite{Winzer2018}, interferometric sensing and optical frequency metrology \cite{Xu2013e}.
Rare-earth ion doping \cite{Polman1991,Polman1997} can equally provide the basis for compact erbium-doped waveguide amplifiers (EDWAs)\cite{Bradley2011}. 
Indeed, pioneering efforts in the 1990s have been made to implement EDWAs based on oxide glass waveguides\cite{Nykolak1993,Hattori1994}. 
Yet, these approaches were limited by large waveguide background losses, large device footprints and incompatibility with contemporary photonic integrated circuits\cite{Thomson2016}, and ultimately abandoned.
Interest in EDWAs re-emerged with the \ce{Si3N4} CMOS-compatible photonic integrated circuit platform, with advantages over silicon including its wider transparency window\cite{Moss2013a}, absence of two-photon absorption in telecommunication bands, a lower temperature sensitivity, high power handling of up to tens of Watts\cite{Brasch2014}, and - most crucially - record low propagation losses of only $<$3~dB/m that can be maintained over meter-scale lengths\cite{Liu2020f}.

One challenge in realizing photonic integrated circuit based erbium amplifiers is the limited gain that can be achieved - stemming from constraints in doping concentration due to cooperative upconversion\cite{Kik2003}.
This limitation necessitates waveguides with low propagation loss and long waveguides with 10s of centimeter to meter lengths, to achieve a large gain and a high output power,  which has until recently been challenging in integrated photonics.
While net (and very significant) gain has been shown, all prior work so far using materials such as erbium-doped \ce{Al2O3} \cite{Mu2020,Ronn2019} and \ce{TeO2}\cite{N2020a} achieved only very limited output powers of typically $<$1~mW.
Despite high doping concentrations, past attempts using atomic layer deposition of \ce{Al2O3} and \ce{Er2O3} layers\cite{Ronn2019} or single-crystal erbium chloride silicate nanowire\cite{Sun2017} only deliver \SI{\ll 1}{\micro\watt} output power.
Such output power is far below the level demanded by many applications, i.e., in the range of 10-100 mW that has been achieved with heterogeneous integration of III-V amplifiers onto silicon photonics\cite{Davenport2016, VanGasse2019, OpdeBeeck2020}.

Here, we overcome these challenges and demonstrate a photonic integrated circuit based erbium-implanted \ce{Si3N4} (\ce{Er\dop Si3N4}) amplifier based on meter-scale ultralow-loss \ce{Si3N4} waveguides.
The integrated \ce{Er\dop Si3N4} amplifier achieves an on-chip output power of 145~mW (21.6~dBm) and a small-signal gain of more than 30~dB.
This result presents a gain performance on par with commercial EDFAs, a more than 100-fold improvement with respect to existing photonic integrated circuit based EDWAs\cite{Mu2020,Ronn2019,N2020a},  and exceeding what has been achieved (17.5 dBm output power) in state-of-the-art heterogeneously integrated III-V amplifiers in silicon-photonics\cite{VanGasse2019} (see Supplementary Information),  and makes the technology viable for real-world applications.

\begin{figure*}[t!]
\centering
\includegraphics[width=\textwidth]{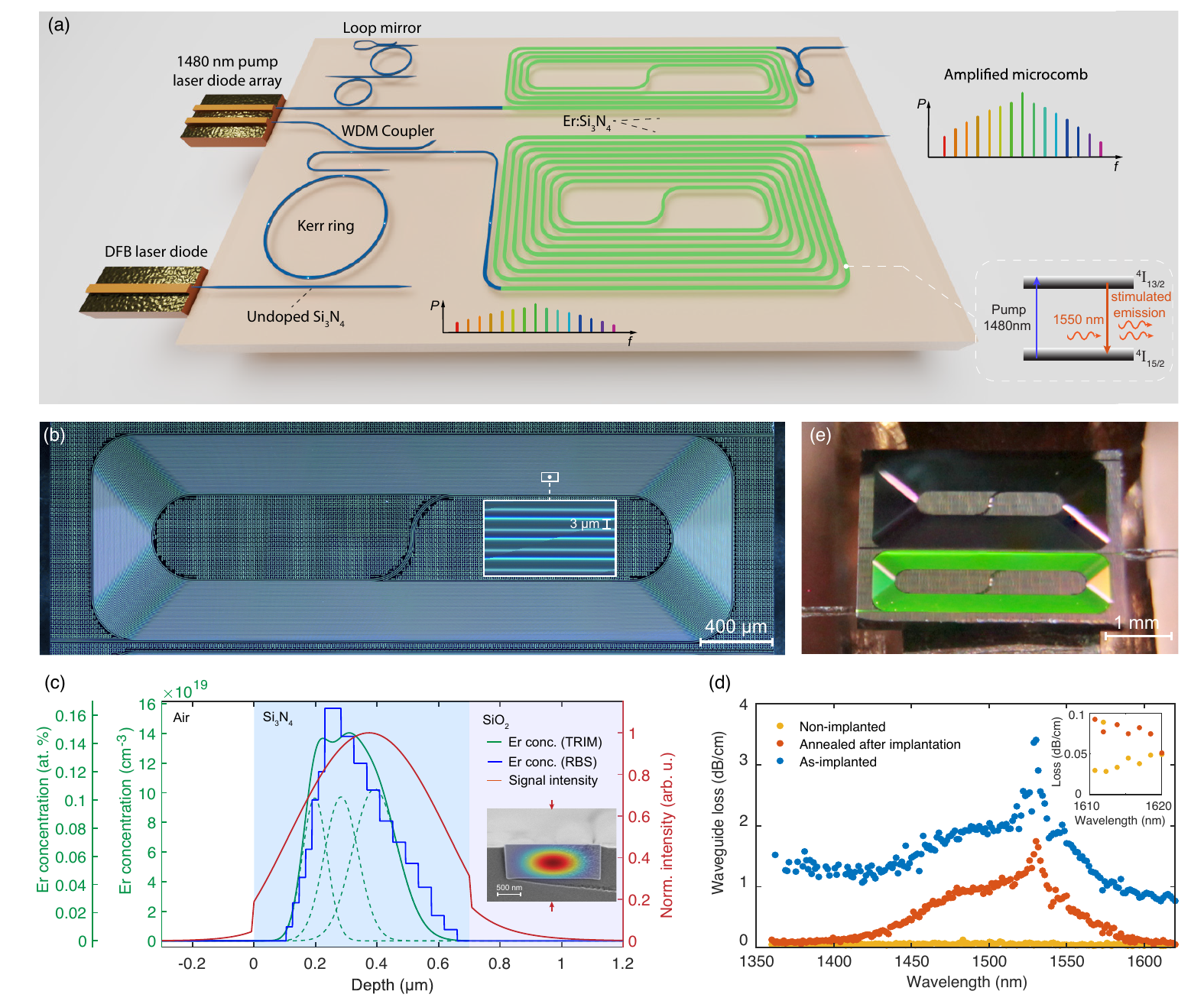}
\caption{
\footnotesize
\textbf{Integrated erbium-implanted \ce{Si3N4} waveguide amplifier.} 
(\textbf{A}) Illustration of an envisaged integrated active \ce{Si3N4} photonic circuit for lasers and optical amplification, leveraging erbium-implanted \ce{Si3N4} waveguides, ultralow-loss passive circuits and Kerr nonlinear devices. The inset shows the optical amplification process of erbium ions excited by the 1480 nm pump. 
(\textbf{B}) Optical image of a 0.5-m-long \ce{Er\dop Si3N4} waveguide coil.
(\textbf{C}) Profiles of the calculated erbium concentration (green) comprising three successive ion implants (dashed green) using SRIM simulations, the simulated optical transverse electric (TE) mode intensity (red) and the measured erbium concentration (blue) by Rutherford backscattering spectrometry (RBS), along the vertical cutline indicated by the arrows in the inset.
The inset shows the SEM image of the waveguide cross section overlaid with the simulated fundamental TE mode.
(\textbf{D}) Measured optical losses of a \ce{Si3N4} waveguide before implantation, as-implanted, and after annealing.
The inset shows the background loss around 1610 nm with weak erbium absorption.
(\textbf{E}) Optical image of the \ce{Er\dop Si3N4} chip with waveguides butt-coupled with two optical fibers. 
The green light emission results from the second-order cooperative upconversion process upon intense optical pumping.
}
\label{Fig:1}
\end{figure*}

In this work, we fabricate up to 0.5 meter long densely-packed \ce{Si3N4} (\fref{Fig:1}(b)) spiral waveguides with a cross section of \SI{0.7 \times 2.1}{\micro\metre\tothe{2}} using the photonic damascene process\cite{Liu2020f}, which exhibit ultralow propagation losses of $<$~5 dB/m.
We do not add an upper oxide cladding to the devices before implantation. 
Next we apply ion implantation\cite{Polman1991}, a wafer-scale process that benefits from much lower cooperative upconversion compared to co-sputtered films\cite{Kik2003},  to the \ce{Si3N4} integrated circuits (see Supplementary Information). 


\begin{figure*}[t!]
\centering
\includegraphics[clip,scale=1]{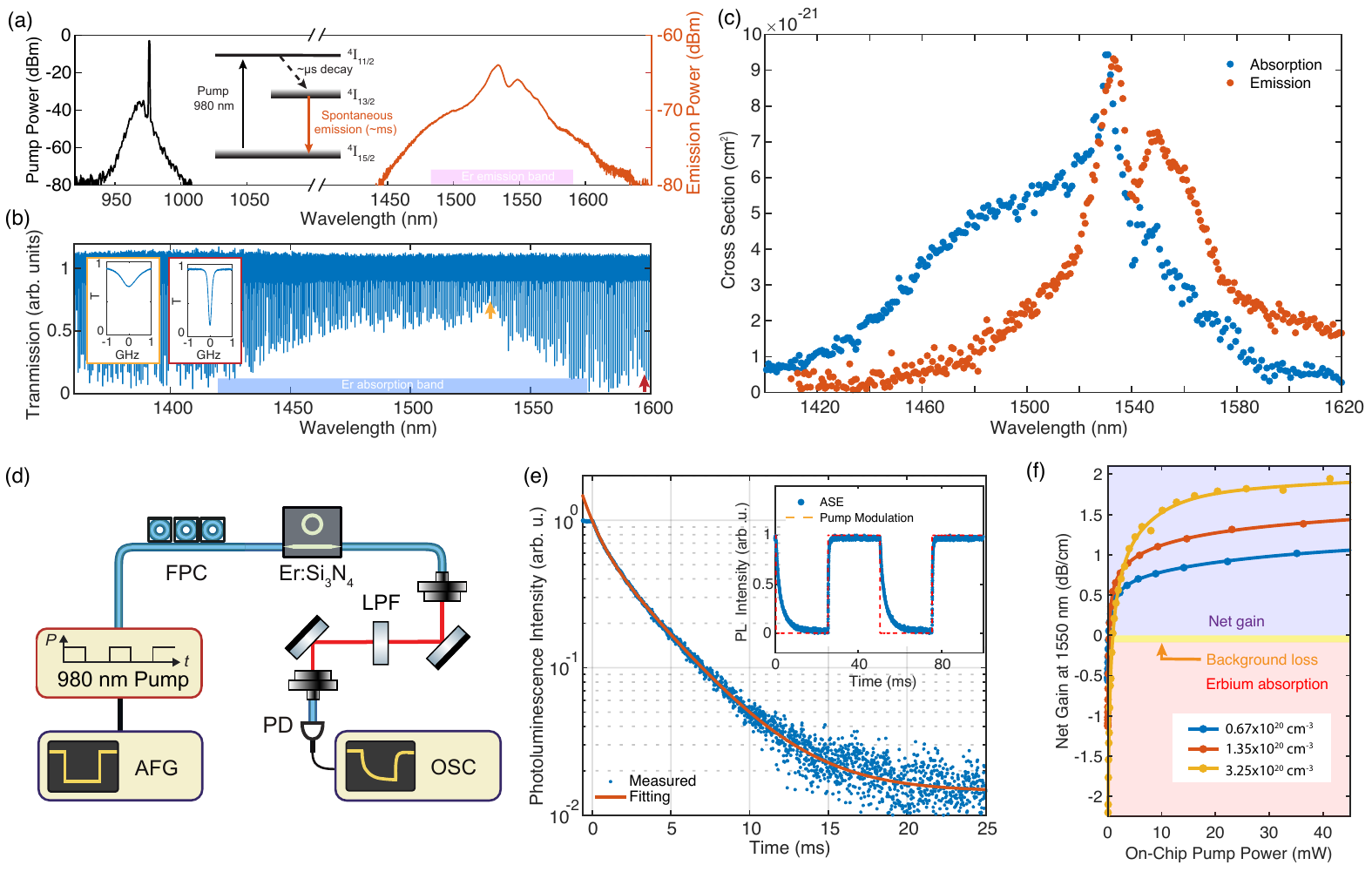}
\caption{
\footnotesize
\textbf{Light emission and absorption properties of \ce{Er\dop Si3N4} photonic waveguides}. 
(\textbf{A}) Fluorescence emission spectrum (in red) of a 0.46-cm-long \ce{Er\dop Si3N4} waveguide upon 980~nm pumping (black), via the emission transition among energy levels illustrated by the inset.
(\textbf{B}) Measured resonance linewidths of an \ce{Er\dop Si3N4} microring resonator used for characterizing the wavelength-dependent erbium absorption.  
The insets exemplify two resonances at the wavelengths indicated by arrows.
(\textbf{C}) Emission and absorption cross sections converted from the measured fluorescence and absorption spectra.  
(\textbf{D}) Experimental setup for fluorescence lifetime measurement.  
A continuous-wave 980 nm laser gated at 20~Hz with 50$\%$  duty cycle is used to pump the \ce{Er\dop Si3N4} waveguide (very weakly coupled with a microring resonator).  
AFG, arbitrary function generator; FPC, fiber-based polarization controller; LPF, optical low pass filter; PD, photodetector; OSC, electrical oscilloscope.
(\textbf{E}) Measured fluorescence decay from the ${}^4\text{I}_{13/2}$ level.  
The inset shows the gated pump and emission. 
(\textbf{F}) Measured net gain per centimeter in the \ce{Er\dop Si3N4} waveguides with different erbium concentrations upon 1480 nm pumping. 
The shaded area in light blue indicates the on-chip net gain regime after excluding the erbium absorption and the waveguide loss. 
}
\label{Fig:2}
\end{figure*}

This endows the \ce{Si3N4} platform with gain, as schematically shown in \fref{Fig:1}(a).  
Despite adopting a 0.5-m-long  \ce{Er\dop Si3N4} waveguide, using spiral coil arrangements with a gap spacing of \SI{3}{\micro\metre} achieves a compact footprint of only $1.2\times3.6$ mm\textsuperscript{2} (\fref{Fig:1}(b)).
In order to achieve a large overlap of  $\Gamma \approx  50\%$ between the embedded ions with the fundamental optical mode of the waveguide, we use three successive implantation steps to the \ce{Si3N4} waveguide using ion energies of 2.0, 1.416 and 0.966 MeV and corresponding ion fluences of 4.50$\times$10$^{15}$, 3.17$\times$10$^{15}$ and 2.34$\times$10$^{15}$~cm$^{-2}$, respectively. 
\Fref{Fig:1}(c) shows the simulated concentration profile (in green) with a maximum depth of 400 nm in \ce{Si3N4} using the Monte-Carlo program package 'Stopping and Range of Ions in Matter' (SRIM), which matches well with Rutherford backscattering spectrometry (RBS) measurement (see Supplementary Information).

Upon implantation, we observe the increase in waveguide background loss from $<$5 dB/m to 100 dB/m due to implantation defects, as shown in \fref{Fig:1}(d).
We observe that the background loss outside the erbium absorption band can be significantly reduced after annealing at \SI{1000}{\degreeCelsius} in oxygen for one hour, approaching the same level as undoped waveguides.  
The loss is extrapolated from intrinsic resonance linewidth measurement of a doped microring resonator (see Supplementary Information). 
Moreover, we observe upon annealing green luminescence stemming from cooperative upconversion when intensively pumping the devices at 1480~nm ( \fref{Fig:1}(e)).
Notably, such a loss only contributes $<$2.5~dB background loss for a 0.5-m-long waveguide, significantly lower than the 30~dB passive loss for waveguides of equal lengths in prior work\cite{Mu2020,Ronn2019}, which deplete the pump early and prevent efficient amplification.
Moreover, we investigate selective masking (using photoresist) during ion implantation, which can provide a basis for integrating both passive and active components in \ce{Si3N4} waveguides, e.g. combining Kerr frequency combs with erbium waveguide amplifiers (see Supplementary Information).
Although waveguide deformations in the implanted regions are observed, the waveguide cross section can be well sustained due to the mechanical support of the lateral oxide cladding (the inset of \fref{Fig:1}(c)), thereby minimizing reflections and enabling low-loss transitions from gain regions to passive waveguide regions ( \fref{Fig:1}(a)) (see Supplementary Information).
In contrast, for the same implantation conditions, severe deformations of \ce{Si3N4} waveguides without cladding are observed (see Supplementary Information).


\begin{figure*}[t!]
\centering
\includegraphics[width=\textwidth]{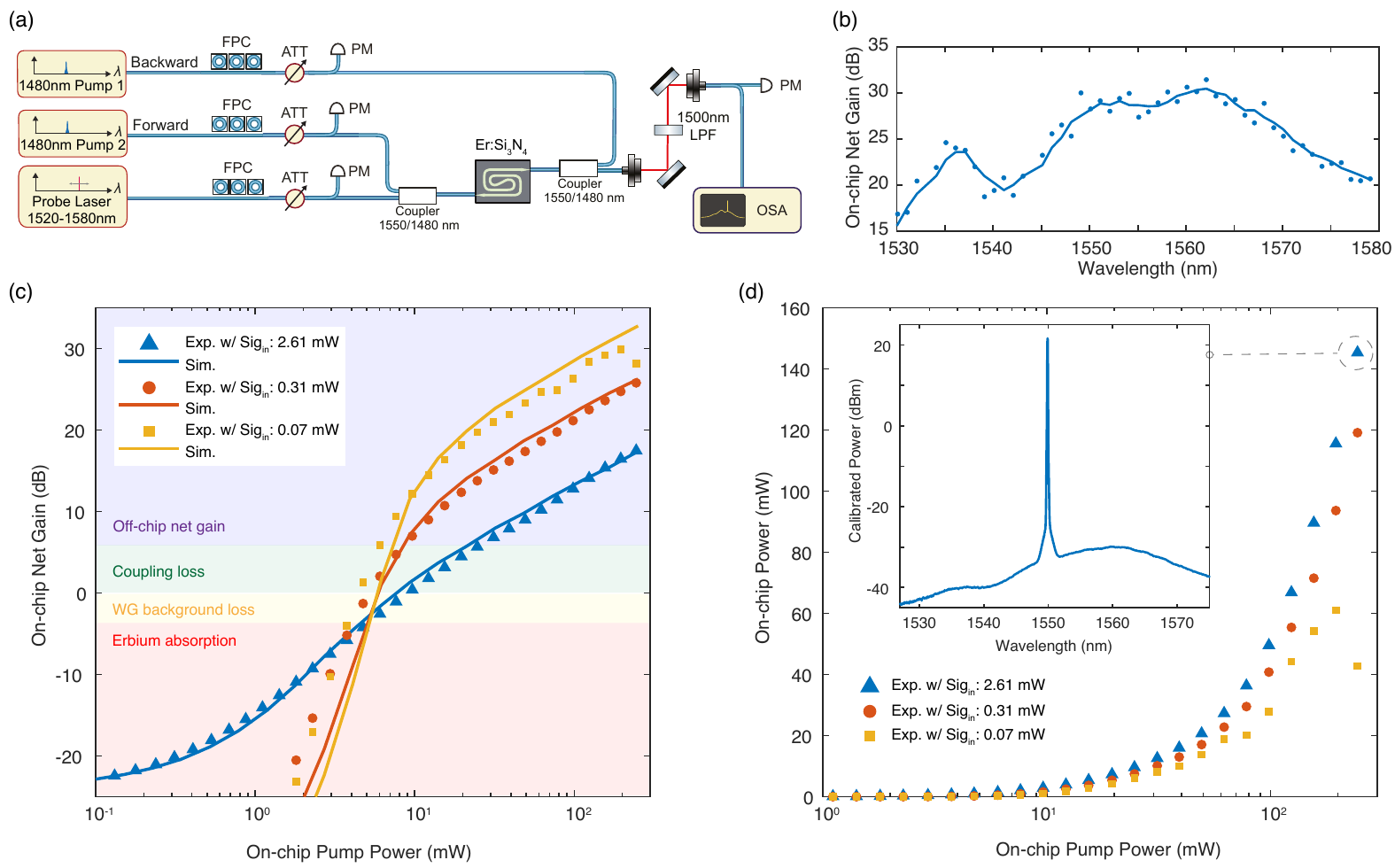}
\caption{
\footnotesize
\textbf{Intense optical amplification in a spiral \ce{Er\dop Si3N4} waveguide amplifier. } 
(\textbf{A}) The experimental setup for measuring the optical amplification.  
A free-space low pass filter is used to isolate the residual pump for amplification analysis. 
FPC, fiber-based polarization controller; ATT, tunable optical attenuators; PM, power meter; LPF, optical low-pass filter; OSA, optical spectrum analyzer.
(\textbf{B}) Measured wavelength-dependent on-chip net gain. 
The solid line indicates the average value of the gain obtained with 0.07~mW on-chip signal input at a total on-chip pump power of 245 mW. 
(\textbf{C}) Measured (scatters) and simulated (solid curves) on-chip gain for signals at 1550 nm. 
The color-shaded areas indicate the regions for off-chip net gain, as well as the loss contributions from fiber-to-chip coupling loss, waveguide background loss and erbium absorption, respectively.
(\textbf{D}) The corresponding on-chip output powers at 1550 nm.  The inset shows the calibrated optical spectrum of the 145~mW signal output after amplification.  
}
\label{Fig:3}
\end{figure*}


\begin{figure*}[t!]
\centering
\includegraphics[width=\textwidth]{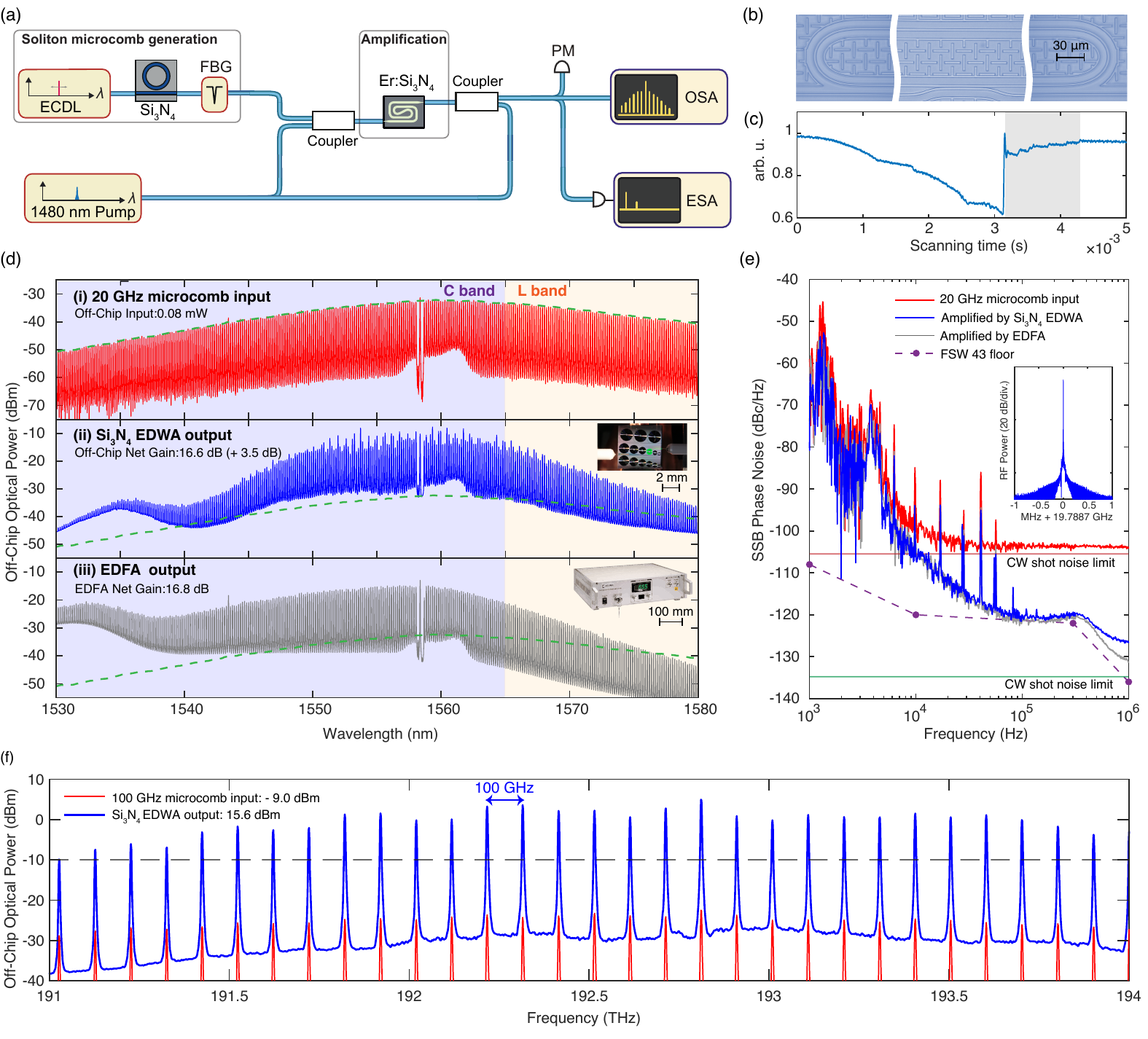}
\caption{
\footnotesize
\textbf{Broadband on-chip amplification of soliton microcombs for low-noise microwave generation and optical communication carriers.} 
(\textbf{A}) The experimental setup for broadband soliton amplification using an integrated \ce{Er\dop Si3N4} waveguide. 
ECDL, external cavity diode laser; FBG, fiber Bragg grating; PM, power meter; OSA, optical spectrum analyzer; ESA, electrical spectrum analyzer.
(\textbf{B}) Stitched optical image of a 19.8~GHz \ce{Si3N4} race-track microring resonator.
(\textbf{C}) The transmission with soliton steps during laser scanning a resonance.
(\textbf{D}) Amplification of 19.8 GHz microcomb in C- and L-band. 
(\textbf{i}) The generated single-soliton state microcomb as the input,  (\textbf{ii}) the amplified microcomb by the the \ce{Er\dop Si3N4} EDWA, and (\textbf{iii}) the amplified microcomb by a commercial EDFA (Calmar Laser AMP-STY).  
The dashed green curves indicate input microcomb envelope. 
The spectral power is calibrated. 
Insets show the photos of the deployed amplifiers.
(\textbf{E}) The corresponding single-sideband (SSB) phase noises of generated microwave signals. The 'step-like' noise feature is due to the analyzer noise floor indicated by the dashed line. 
(\textbf{F}) The optical spectra of amplified 100~GHz soliton microcomb (36~mW) and the input (0.12~mW). The corresponding 24.8 dB net gain allows for achieving a maximum comb line power of $>$ 1 mW in the optical communication C-band. 
}
\label{Fig:4}
\end{figure*}

The emission and absorption cross-sections ($\sigma_\mathrm{e}(\lambda)$ and $\sigma_\mathrm{a}(\lambda)$) are important parameters 
that can determine the erbium gain property. 
We next infer $\sigma_\mathrm{e}(\lambda)$ and $\sigma_\mathrm{a}(\lambda)$  by examining the photoluminescence (PL) spectrum and the wavelength($\lambda$)-dependent absorption loss,  respectively.  
\Fref{Fig:2}(a) shows the emission spectrum of the transition from the first excited level to the ground state by resonantly pumping the ${}^4{I}_{11/2}$ state (980 nm) in a 0.46-cm-long \ce{Er\dop Si3N4} waveguide, from which the emission cross-section ( \fref{Fig:2}(c)) can be derived via\cite{McCumber1964}
\begin{equation}
 \frac{1}{\tau}=8 \pi n^{2} c \int \frac{\sigma_{\mathrm{em}}(\lambda)}{\lambda^{4}} \mathrm{~d} \lambda,
 \label{eqn:emission}
\end{equation}
 where $n$ is the refractive index and $\tau=3.4$~ms is the PL lifetime extracted from the temporal measurement as shown in \fref{Fig:2}(d) and (e).  The wavelength-dependent erbium absorption loss $\alpha_{Er}$~(in cm$^{-1}$) is derived from the intrinsic resonance linewidths of an erbium-doped microring resonator.
The absorption is subsequently converted to the erbium absorption cross-section using
\begin{equation}
\sigma_{a}(\lambda)=\frac{\alpha_{\mathrm{Er}}(\lambda)}{ \Gamma \cdot N_{0}},
\label{eqn:absorption}
\end{equation}
where $N_0$ is the effective peak erbium ion concentration and $\Gamma$ is the overlap factor.  
The insets of \fref{Fig:2}(b) show two resonances with intrinsic dissipation rates ($\kappa_0/2\pi$) of $\sim$1000~MHz and $\sim$40~MHz at 1535 nm and 1600 nm, corresponding to loss coefficients of 2 dB/cm and 0.08 dB/cm, respectively.   
The measured losses are calibrated by optical frequency domain reflectometry (OFDR) measurements (see Supplementary Information).
The absorption cross section $\sigma_{a}(\lambda)$ is obtained by scaling its peak value to that of the derived $\sigma_{em}(\lambda)$, which in turn gives the $N_0$ of $1.35\times 10^{20}~\text{cm}^{-3}$ .
Compared to the RBS result, almost all the incorporated erbium ions are optically active.
Different erbium concentrations in \ce{Er\dop Si3N4} waveguides can be achieved by scaling the ion fluences.
In reference \ce{Er\dop Si3N4} waveguides with ion concentration $N_0$ of $0.67\times 10^{20}~\text{cm}^{-3}$, $1.35\times 10^{20}~\text{cm}^{-3}$ and $3.25\times 10^{20}\text{cm}^{-3}$, we obtained net gain coefficients of 1.0~dB/cm, 1.4~dB/cm and 1.9~dB/cm at 1550~nm, respectively (\fref{Fig:2}(f)), comparable to the theoretical gain coefficients of 0.7~dB/cm, 1.5~dB/cm and 3.5~dB/cm given by $g=\Gamma(\sigma_\mathrm{e}N_2 - \sigma_\mathrm{e}N_1)$ where $N_2$ and $N_1$ are the populations of the first excited state and the ground state\cite{Giles1991}.
The gain coefficient with 980 nm pump and its dependence on waveguide width is also investigated (see Supplementary Information).

Next, we demonstrate a high output power and a large net gain upon 1480~nm pumping in a 0.21-meter-long \ce{Er\dop Si3N4} waveguide with an erbium concentration of ca. \SI{3.25\times 10^{20}}{\centi\metre\tothe{-3}}, using the setup as shown in \fref{Fig:3}(a).
The broadband on-chip net gain reaches 30 dB over the wavelength range from 1550 nm to 1565 nm (\fref{Fig:3}(b)).
\Fref{Fig:3}(c) shows measured on-chip optical net gain at 1550 nm, along with simulations reproduced by parameter fitting.
The on-chip output power reaches 60 mW for 0.07~mW input, yielding a maximum net gain of 30~dB and a power conversion efficiency of about $30\%$ (signal power increment divided by pump power).  
This indicates an off-chip net gain of 24 dB when considering a two-side fiber-to-chip coupling loss of 5.8 dB. 
We achieve an on-chip output power reaching 145~mW for an increased input power of 2.61~mW at 245~mW coupled pump power,  as shown in \fref{Fig:3}(d) and the inset, indicating an on-chip power conversion efficiency approaching $60\%$.
A similar gain performance is achieved in a 0.5-meter-long \ce{Er\dop Si3N4} waveguide with a lower erbium concentration of ca.  \SI{1.35\times 10^{20}}{\centi\metre\tothe{-3}} (see Supplementary Information).
A noise figure of ca. 7 dB is measured at net gain of $>$20~dB, limited by coupling losses (see Supplementary Information).
Simulations suggest that a higher net gain and output power can be achieved with a higher pump power, without significantly suffering from the cooperative upconversion that can fundamentally limit the performance of EDWAs using co-sputtered erbium-doped films\cite{Kik2003} (see Supplementary Information).
It should be noted that the observed fluctuations in the measured net gain (yellow scatters) are caused by the gain clamping effect once the net gain reaches about 26~dB,  which originates from the competing parasitic lasing in an optical Fabry-P\'erot (FP) cavity formed by two waveguide facets (see Supplementary Information).

As an example of the utility, we apply the \ce{Er\dop Si3N4} EDWA to soliton microcomb amplification (\fref{Fig:4}(a)) which, given both devices use \ce{Si3N4},  can in principle be integrated on the same chip.
Soliton microcombs can provide fully coherent broadband optical frequency combs when operating in the dissipative Kerr soliton (DKS) regime\cite{Summary2018}.
Yet, soliton microcombs exhibit a low intrinsic nonlinear conversion efficiency (e.g. , $<1\%$ for a 200 GHz spacing single-soliton microcomb\cite{Bao2014a}), requiring bench-top EDFAs to be used in virtually all applications. 
Here, we demonstrate on-chip amplification of soliton microcombs to 10 mW level across the C- and L-band, suitable for photonic generation of microwaves and in wavelength-division-multiplexing (WDM) coherent optical communications.
First, a 19.8~GHz single-soliton microcomb is generated  in a \ce{Si3N4} Euler-bend racetrack microresonator (\fref{Fig:4}(b) and (c)) with 0.08~mW output power, as shown by panel (i) in \fref{Fig:4}(d). 
The amplified soliton with 8.4~mW off-chip power by the \ce{Er\dop Si3N4} EDWA (panel (ii)) leads to a significant reduction in single-sideband phase noise to $<$-120~dBc/Hz at Fourier offset frequencies at $>100$~kHz, compared to -104~dBc/Hz (before amplification) that is limited by photon shot noise, as shown in \fref{Fig:4}(e).
To provide a comparison, a commercial EDFA is also deployed to amplify the same soliton microcomb to a similar power level (panel (iii) in \fref{Fig:4}(d)), giving identical performance. 
As a second example, we amplify a soliton microcomb with a channel spacing of 100 GHz suitable for WDM optical communications. 
We observe more than 30 comb lines achieving line powers of $>0.1$ mW (-10 dBm) with an optical signal-to-noise ratio (OSNR) exceeding 30 dB (0.05 nm resolution bandwidth), among which more than 10 comb lines exceed 1 mW (0 dBm), as shown in \fref{Fig:4}(f).
The observed comb line OSNR is envisaged to feasibly satisfy the requirement of the soliton-based transmitter for short-reach coherent data transmission at a symbol rate of 40~gigabauds per channel using 16-state quadrature amplitude modulation based on the same configurations as past work \cite{Marin-Palomo2017} (see Supplementary Information).


In summary, we have demonstrated a photonic integrated circuit based \ce{Er\dop Si3N4} waveguide amplifier that can provide up to 145~mW on-chip output power and a small-signal gain of more than 30~dB. 
The design freedom afforded by photonic integrated circuits allows multi-stage configurations to be adopted, in order to optimize the gain and the optical signal-to-noise ratio.  
Moreover, the ion implantation technique could allow for co-doping other rare-earth ions such as ytterbium (emission at \SI{1.1}{\micro\metre}) and thulium (\SI{0.8}{\micro\metre},  \SI{1.45}{\micro\metre} and \SI{2.0}{\micro\metre}), thereby providing gain in other wavelength regions.
The presented technique of ion implantation in \ce{Si3N4} can serve as the gain medium in  a variety of integrated laser sources such as high-power soliton microcombs, low-noise rare-earth-ion-based CW lasers, femtosecond mode-locked lasers\cite{Singh2020}  or cavity soliton lasers\cite{Bao2019}.  
Equally important, this active \ce{Si3N4} photonic platform is compatible with heterogeneous integration of thin-film lithium niobate\cite{Churaev2021,Snigirev2021}, enabling the combination of both high speed electro-optical modulation and amplification on the same chip, of use for coherent communications\cite{Marin-Palomo2017a} or radio-frequency distribution\cite{Marpaung2019}.

\medskip
\begin{footnotesize}
\noindent \textbf{Funding Information}: This work was supported by Contract HR0011-20-2-0046 (NOVEL) from the Defense Advanced Research Projects Agency (DARPA), by the Air Force Office of Scientific Research (AFOSR) under Award No. FA9550-19-1-0250, by the Swiss National Science Foundation under grant agreement No. 176563 (BRIDGE) and grant no. 201923 (Ambizione), and by the collaborative research center CRC 1375 NOA (Project C5). This work is also funded by the EU H2020 research and innovation programme under grant No. 696656 (Graphene Core3), grant No. 965124 (FEMTOCHIP) and Marie Sklodowska-Curie IF grant No. 898594 (CompADC).  

\noindent \textbf{Acknowledgments}: 
Silicon nitride samples were fabricated in the EPFL Center of MicroNanoTechnology (CMi).  
We thank Arslan S. Raja, Wenle Weng and Tianyi Liu for discussions and assistance.
We thank Prof. Elke Wendler for her help in analyzing the RBS data.  

\noindent \textbf{Author contributions}: 
Y.L.  and Z.Q.  performed the experiments, data analysis and simulations. 
X.J.,  J.H.  and J.R. provided experimental supports. 
J.L., R.N.W., Z.Q.  and X.J.  designed and fabricated the passive \ce{Si3N4} samples. 
Y.L.  and Z.Q.  designed \ce{Si3N4} waveguide amplifier samples.
Z.Q.  performed chip masking.
M.H.  and C.R.  performed the RBS measurement and analysis. 
Y.L.  wrote the manuscript with the assistance from Z.Q.  and the input from all co-authors. 
T.J.K supervised the project.

\noindent \textbf{Data Availability Statement}: 
The code and data used to produce the plots within this work will be released on the repository \texttt{Zenodo} upon publication of this preprint.

\noindent \textbf{Competing interests}
T.J.K. is a cofounder and shareholder of LiGenTec SA, a start-up company offering \ce{Si3N4} photonic integrated circuits as a foundry service. 

\noindent \textbf{Supplementary Materials} Materials and Methods;\\
 Supplementary Text; Figs. S1 to S12; Tables S1 to S3;\\
 References \textit{(33-76)}.

\end{footnotesize}

\renewcommand{\bibpreamble}{
$\ast$These authors contributed equally to this work.\\
$\dagger$\textcolor{magenta}{yang.lau@epfl.ch}\\
$\ddag$\textcolor{magenta}{tobias.kippenberg@epfl.ch}
}
\pretolerance=0
\bigskip
\bibliographystyle{apsrev4-2}
\bibliography{library}
\end{document}


\title{Supplementary Information for: A photonic integrated circuit based erbium-doped amplifier}

\author{Yang Liu$^{1,\ast,\dagger}$, 
				Zheru Qiu$^{1,\ast}$,
			    Xinru Ji$^{1}$, 
			    Jijun He$^{1}$, 
			    Johann Riemensberger$^{1}$, 
			    Martin Hafermann$^{2}$,
			    Rui Ning Wang$^{1}$,
			    Junqiu Liu$^{1}$,
				Carsten Ronning$^{2}$,
				and Tobias J.  Kippenberg$^{1,\ddag}$}
				
\affiliation{
$^1$Institute of Physics, Swiss Federal Institute of Technology Lausanne (EPFL), CH-1015 Lausanne, Switzerland\\
$^2$Institute of Solid State Physics, Friedrich Schiller University Jena, Max-Wien-Platz 1, 07743 Jena, Germany
}

\setcounter{equation}{0}
\setcounter{figure}{0}
\setcounter{table}{0}

\setcounter{subsection}{0}
\setcounter{section}{0}

\begin{abstract}
Supplementary Information accompanying the manuscript containing performance comparison with prior works, sample fabrication process,  simulations of ion doping, theoretical analysis of optical gain and characteristic parameters, experimental details including measurement methods, calibrations and  characterizations.
\end{abstract}

\maketitle
{\hypersetup{linkcolor=black}\tableofcontents}
\newpage


\section{Gain performance comparison with state-of-the-art on-chip amplifiers.}
Table \ref{tab:comparison} summarizes the state-of-the-art prior works on integrated optical amplifiers, including planar amplifiers based on erbium-doped waveguide core or coating, as well as heterogeneously integrated III-V group semiconductor amplifiers.
We compare the \ce{Er\dop Si3N4} waveguide amplifier demonstrated in this work with the prior works and few commercial general purpose erbium-doped fiber amplifiers in terms of achievable on-chip output optical power, net gain, etc.
The comparison shows that our PIC-based \ce{Er\dop Si3N4} amplifier achieves a significant improvement in the output power and gain over reported erbium-doped approaches, reaching a comparable performance with the state-of-art heterogeneously integrated III-V amplifiers.
We also noted some amplification performance characteristics demonstrated here are on par with or even better than the specifications of some commercial erbium-doped fiber amplifiers. We can imagine the demonstrated PIC-based \ce{Er\dop Si3N4} amplifier can replace such amplifiers in various applications and the small size of gain section can result in a significant form factor reduction.

\begin{table*}[h!] 
\caption{\textbf{Comparison with reported on-chip amplifiers and typical commercial erbium doped fiber amplifiers.}
The gain performance of the demonstrated \ce{Er\dop Si3N4} waveguide amplifier is compared with the state-of-the-art on-chip amplifiers based on erbium-doped gain media and heterogeneously integrated III-V group semiconductors.
Three typical commercially available general purpose erbium-doped fiber amplifiers are also compared. n/a, data not available.
}
\label{tab:comparison}
\centering
\bgroup 
\def\arraystretch{1.2}
\begin{tabular}{llcccll}
	\toprule
	\thead{Waveguide\\ material}           & \thead{Active\\ material}              & \thead{Maximum\\ on-chip output\\ power (dBm)} & \thead{Peak on-chip\\ net gain\\ (dB)}  & \thead{Length (cm)} & \thead{Erbium \\ concentration \\ ($\text{cm}^{-3}$)} & \thead{Reference}                                    \\ \hline
	                                                                                                                \multicolumn{7}{c}{Si and \ce{Si3N4} photonic integrated circuit based devices.}                                                                                                                 \\ \hline
	\textbf{\ce{Er\dop Si3N4}}             & \textbf{\ce{Er\dop Si3N4}}             &                 \textbf{21.6}                  &               \textbf{30}               &     \textbf{21}     & $\mathbf{3.25\times10^{20}}$                          & \textbf{This Work}                                   \\ 
	\hline
	\ce{Si3N4}/\ce{Er\dop Al2O3}           & \ce{Er\dop Al2O3}                      &                     $-3.5$                     &                   8.5                   &         5.9         & $1.7\times10^{20}$                                    & Chrysostomidis et al.\cite{Chrysostomidis2021}       \\ 
	\ce{Si3N4}/\ce{Er\dop Al2O3}           & \ce{Er\dop Al2O3}                      &                      2.7                       &                  18.1                   &         10          & $1.65\times10^{20}$                                   & Mu et al.\cite{Mu2020}                               \\ 
	\ce{Si3N4}                             & \ce{Er\dop TeO2}                       &                     $< -1$                     &                    5                    &         6.7         & $2.5\times10^{20}$                                    & Frankis et al.\cite{N2020a}                          \\ 
	\ce{Si3N4}                             & \ce{Er\dop Al2O3}                      &                    $< -20$                     &                   0.4                   &        0.16         & $1.11- 3.88\times10^{21}$                             & R\"onn et al.\cite{Ronn2019}                         \\ 
	\ce{Si}                                & \ce{Er\dop Al2O3}                      &                     $<-4$                      &                  1.86                   &        0.95         & $2.7\times10^{20}$                                    & Agazzi et al.\cite{Agazzi2010}                       \\ 
\hline
	                                                                                                                      \multicolumn{7}{c}{Planar waveguide devices based on oxide materials.}                                                                                                                       \\ \hline
	\ce{Er\dop Yb\dop Al2O3}               & \ce{Er\dop Yb\dop Al2O3}               &                    $<–8.5$                     &                   4.3                   &          3          & $1.5\times10^{20}$                                    & Bonneville \textit{et al.}\cite{Bonneville2020}               \\ 
	\ce{Er\dop Al2O3}                      & \ce{Er\dop Al2O3}                      &                       4                        &                   20                    &        12.9         & $1.92\times10^{20}$                                   & V\'azquez-C\'ordova \textit{et al.}\cite{Vazquez-Cordova2014} \\ 
		\ce{Er\dop Al2O3}                      & \ce{Er\dop Al2O3}                      &                       $<$-37                        &                   2.3                    &        4         & $2.7\times10^{20}$                                   & G. N. van den Hoven \textit{et al.}\cite{VanDenHoven1996} \\
	\ce{Er\dop Ta2O5}                      & \ce{Er\dop Ta2O5}                      &                    $< -20$                     &                  4.83                   &         2.3         & $2.7\times 10^{20}$                                   & Subramanian \textit{et al.}\cite{Subramanian2012}             \\ 
	\ce{Er\dop TeO2}                       & \ce{Er\dop TeO2}                       &                       13                       &                   14                    &          5          & $2.2\times 10^{20}$                                   & Vu and Madden \cite{Vu2010}         \\ 
	\ce{Er\dop LiNbO3}                     & \ce{Er\dop LiNbO3}                     &                       0                        &                   18                    &         3.6         & $1.9\times10^{20}$                                    & Zhou \textit{et al.}\cite{Zhou2021}                           \\ 
\hline
	                                                                                                                              \multicolumn{7}{c}{Low confinement waveguide based.}                                                                                                                               \\ \hline
	\makecell{Er-doped\\ phosphate glass}  & \makecell{Er-doped\\ phosphate glass}  &                      4.4                       &                   27                    &        47.7         & $\approx 4\times10^{19}$                              & Hattori \textit{et al.}\cite{Hattori1994}                     \\ 
	\makecell{Er-doped\\ phosphate glass}  & \makecell{Er-doped\\ phosphate glass}  &                       12                       &                   20                    &         4.1         & $\approx 2.2\times10^{20}$                            & Barbier \textit{et al.}\cite{BARBIER1995}                     \\ 
	\makecell{Er-doped\\ soda-lime glass}  & \makecell{Er-doped\\ soda-lime glass}  &                      5.2                       &                   15                    &         4.5         & $\approx 4\times10^{20}$                              & Nykolak \textit{et al.}\cite{Nykolak1993}                     \\ 
	\makecell{Er-doped\\ soda-lime glass}  & \makecell{Er-doped\\ soda-lime glass}  &                  $\approx 19$                  &                  16.7                   &         3.1         & $\approx 1.8\times10^{20}$                            & Della Valle \textit{et al.}\cite{Della2006a}           \\ 
	\makecell{Er-doped\\ bismuthate glass} & \makecell{Er-doped\\ bismuthate glass} &                  $\approx 16$                  &              $\approx 18$               &         8.7         & $\approx 5.6\times10^{19}$                            & Thomson \textit{et al.}\cite{Thomson2010a}                    \\ 
\hline
	                                                                                                                        \multicolumn{7}{c}{III-V  heterogeneous integrated amplifiers.}                                                                                                                          \\ \hline
	\ce{Si}                                & III-V  material                        &                      17.5                      &                   27                    &        0.145        & n/a                                                   & Van Gasse \textit{et al.}\cite{VanGasse2019}                  \\ 
	\ce{Si}                                & III-V  material                        &                       14                       &                   25                    &         n/a         & n/a                                                   & Davenport \textit{et al.}\cite{Davenport2016}                 \\ 
	\ce{Si3N4}                             & III-V  material                        &                      8.8                       &                   14                    &        0.115        & n/a                                                   & Op de Beeck \textit{et al.}\cite{OpdeBeeck2020}               \\ 
	\ce{LiNbO3}                            & III-V  material                        &                      $<5$                      &                  11.8                   &         n/a         & n/a                                                   & Op de Beeck \textit{et al.}\cite{Eeck2021}                    \\ 
\hline
	                                                                                                                          \multicolumn{7}{c}{Commercial erbium-doped fiber amplifiers (EDFAs).}                                                                                                                           \\ \hline
	Er-doped fiber                         & Er-doped fiber                         &                       18                       &                 25--35                  &         n/a         & n/a                                                   & Calmar Laser AMP-PM-18                               \\
	Er-doped fiber                         & Er-doped fiber                         &                       13                       &                   $>$19                   &         n/a         & n/a                                                   & Amonics AEDFA-BO-13                                  \\
	Er-doped fiber                         & Er-doped fiber                         &                      $>$20                       &                   $>$30                   &         n/a         & n/a                                                   & Thorlabs EDFA100S                                    \\ 
\hline
\end{tabular}
\egroup
\end{table*}

\begin{figure*}[ht!]
	\centering
	\includegraphics[width=\textwidth]{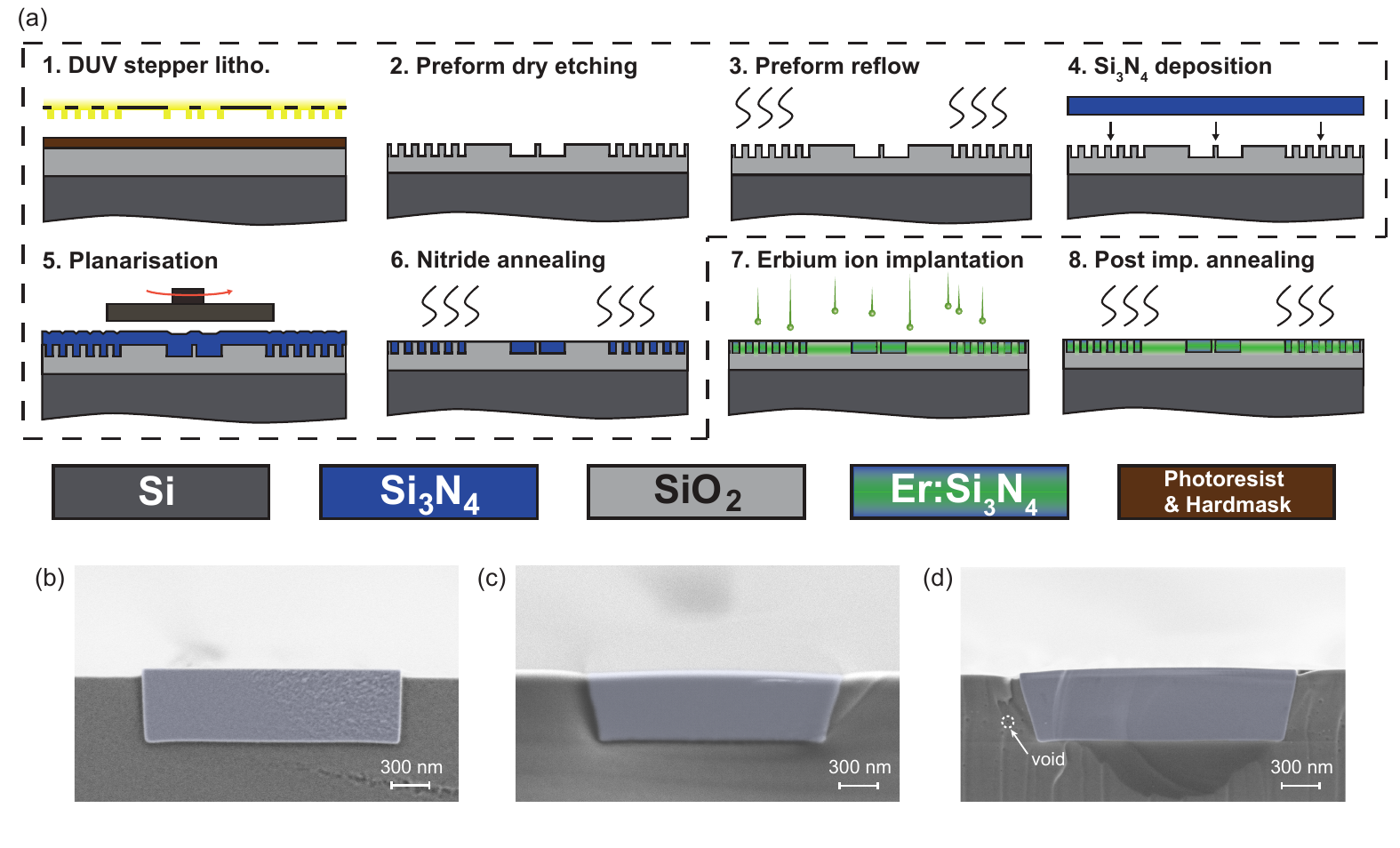}
	\caption{\textbf{\ce{Er\dop Si3N4} photonic chip fabrication}.
		\textbf{a}, The \ce{Er\dop Si3N4} PIC fabrication process consisting of the \ce{Si3N4} photonic Damascene process (dashed line box), the ion implantation and the erbium activation.
		\textbf{b}, False colored SEM image of a \ce{Si3N4} waveguide cross section produced after Step 6, prior to the ion implantation and annealing.
		\textbf{c}, False colored SEM image of a \ce{Si3N4} waveguide produced after Step 8, i.e. post implantation annealing for erbium activation and defect recovery at \SI{1000}{\degreeCelsius} .
		\textbf{d}, False colored SEM image of a \ce{Si3N4} waveguide after modified post implantation annealing at \SI{1200}{\degreeCelsius}.
		Cross section samples are briefly treated in buffered \ce{HF} for $\sim10$~s to create topography contrast under SEM.
		}
	\label{Fig:SIProc}
\end{figure*}

\section{Fabrication process of integrated \ce{Er\dop Si3N4} photonic circuits.}

The fabrication process of the erbium-implanted \ce{Si3N4} PICs\cite{Pfeiffer2016a,Pfeiffer2018,Liu2020f} starts with commercial 4 inch silicon wafers with 4 um thick wet thermal oxide as shown in Fig.\ref{Fig:SIProc}.
A stack of silicon oxide and silicon nitride thin films, or alternatively a layer of amorphous silicon are deposited by low-pressure chemical vapor deposition (LPCVD) to serve as hard-mask.
The waveguide structure and filler pattern for stress release are then defined by deep ultra-violet photolithography (ASML PAS 5500/350C stepper, JSR M108Y resist, and Brewer DUV-42P coating) and transferred into the thermal oxide layer by multiple steps of fluorine chemistry based reactive ion etching (RIE).
Thus, the preform with recesses for the waveguides and filler pattern is created.
In the case of wafers using amorphous silicon as hard-mask, the remaining hard-mask material is stripped in concentrated \ce{KOH} solution at \SI{60}{\degreeCelsius}.
A thermal treatment at \SI{1250}{\degreeCelsius} is then applied to reflow the silicon oxide, reducing the roughness caused by the RIE etching \cite{Pfeiffer2018}, before the preform recesses are filled with stoichiometric \ce{Si3N4} by LPCVD.
Immediately after the deposition, an etchback process consisting of photoresist spin-coating and RIE is performed to roughly planarize the surface and remove most of the excess \ce{Si3N4} material.
Chemical mechanical polishing (CMP) is then applied to reach the desired waveguide thickness, and create a top surface with sub-nanometer root-mean-square roughness.
The wafers are then annealed at \SI{1200}{\degreeCelsius} to drive out hydrogen, which is known to induce light absorption losses at the technologically relevant wavelengths.
Using this process, we create \ce{Si3N4} waveguides buried in a wet oxide cladding but with the top surface exposed, allowing for direct erbium implantation into the waveguides.
We then separate the PIC dies by ultraviolet (UV) photolithography, deep RIE, and backside grinding.
This die separation process creates smooth chip facets which enables high coupling efficiency to optical fibers.
Erbium ion implantation of the \ce{Si3N4} PICs is performed via facilities at the University of Surrey Ion Beam Centre.
For proof-of-concept demonstrations, the separated dies are then attached to silicon carriers up to 2-inch wafer size with adhesives and irradiated in batches.
The samples demonstrating high amplification performance are annealed in a microelectronic grade furnace at \SI{1000}{\degreeCelsius} in \ce{O2} under atmosphere pressure for 1 hour.
We note that it is straightforward to downsize and implant the entire PIC wafer for future large-scale fabrication, which does not require any new technology or even equipment upgrade.
A \ce{SiO2} passivation layer may be optionally applied on top afterwards to enable co-integration of other functional devices such as metal microheaters and piezo actuators.

Fig.~\ref{Fig:SIProc} (b), (c) and (d) compare the scanning electron microscopy (SEM) images of cross sections of reference waveguides on the same wafer before ion implantation, after implantation and annealing at \SI{1000}{\degreeCelsius}, and after implantation and annealing at \SI{1200}{\degreeCelsius}, respectively.  Compared to the waveguide before implantation,  implanted waveguides shown in Fig.~\ref{Fig:SIProc} (c) and (d) exhibit a noticeable waveguide geometry deviation, which could be attributed to the volume expansion of \ce{Si3N4} during ion implantation.
Despite the apparent distortion of the waveguide cross section, according to the ring resonator measurements and OFDR measurements, the effect on the waveguide background loss is still minimal.

In the search for optimum post-implantation annealing process condition, we pushed the annealing temperature to \SI{1200}{\degreeCelsius} for the sample shown in Fig.~\ref{Fig:SIProc} (d). 
We observe voids with a figure size of 10--30~nm after short time buffered \ce{HF} treatment at the lateral sides of the \ce{Si3N4} waveguide rather than underneath the waveguide. 
Such voids indicate the formation of erbium oxide precipitates, which are soluble in \ce{HF}, in the \ce{SiO2} cladding after annealing at more than \SI{1000}{\degreeCelsius}\cite{Polman1991}.
It's reported that the erbium ions in such precipitations are inactive for optical amplification and may lead to elevated unsatuable loss\cite{Prtljaga2012a}. 
In contrast,  Fig.~\ref{Fig:SIProc} (c) shows that the sample annealed at \SI{1000}{\degreeCelsius} does not feature these voids associated with precipitations, which agrees with the observation in \cite{Polman1991}.
Interestingly, such voids are not observed in the \ce{Si3N4} waveguide cores in any sample under inspection.
This provides an evidence showing that unlike \ce{SiO2}, \ce{Si3N4} can be a good host for the erbium dopant, which can accommodate at least $0.3~\mathrm{at.}~\%$ of erbium atoms without precipitation problem in our demonstrations.

\section{Implantation to \ce{Si3N4} waveguides without lateral cladding.}
\begin{figure}[h!]
	\centering
	\includegraphics[width=\textwidth]{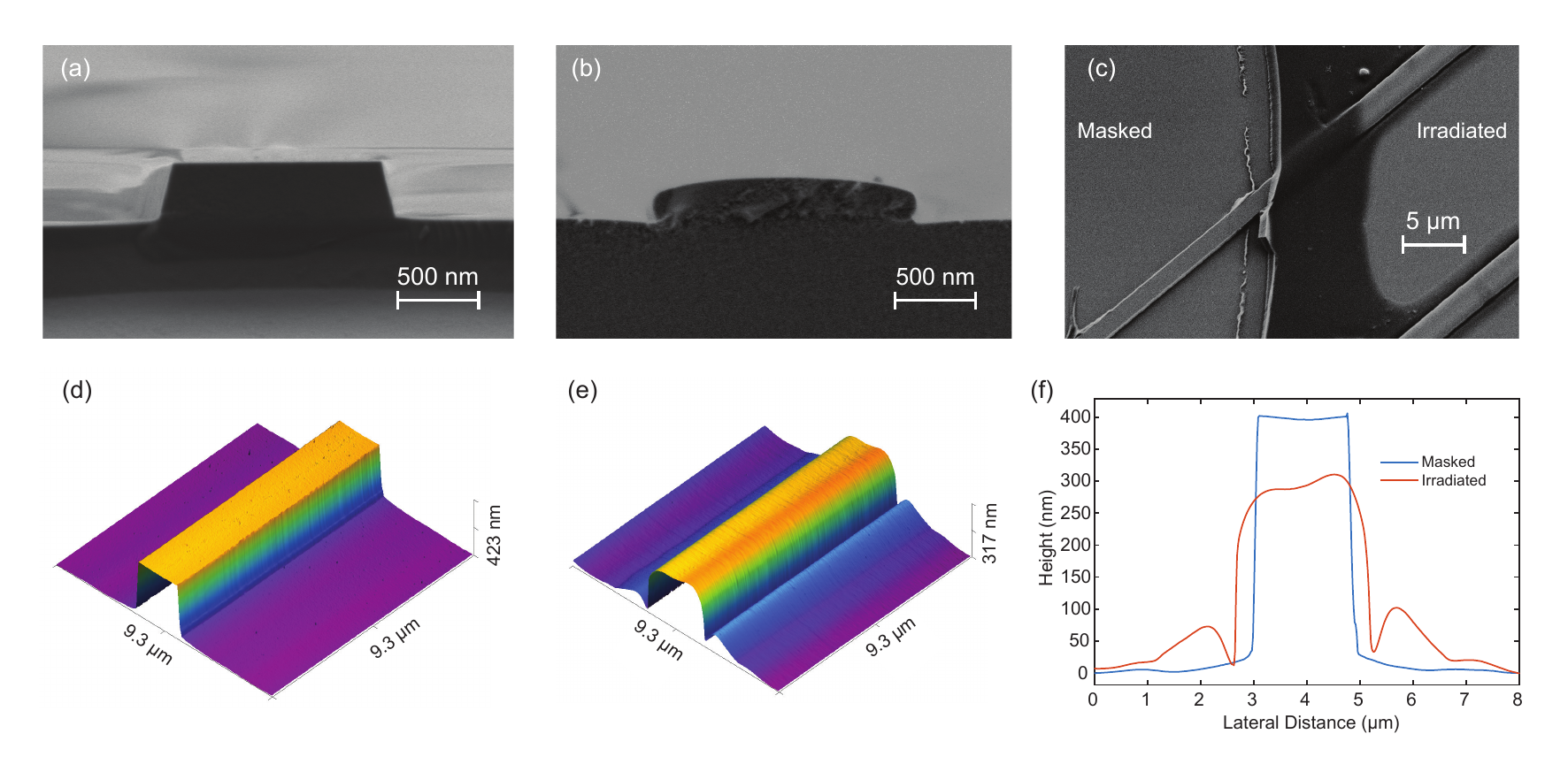}
	\caption{\textbf{Geometry characterizations of \ce{Si3N4} waveguides without lateral cladding before and after erbium implantation.}
		\textbf{a},  SEM image of the \ce{Si3N4} ridge waveguide cross section before ion implantation.
		\textbf{b}, SEM image of the \ce{Si3N4} ridge waveguide cross section after irradiation.
		\textbf{c}, Tilted top-down view SEM image of the boundary region between the irradiated waveguide segment and the non-irradiated segment masked by a droplet of wax (Quickstick 135).
		The masking wax was only partially removed in heated acetone after irradiation.
		\textbf{d}, Surface topography of (a) measured by atomic force microscopy (AFM).
		\textbf{e}, Surface topography of (b) measured by AFM.
		\textbf{f}, Line profile of (d) and (e) in a plane normal to the waveguide axis.
	}
	\label{Fig:SubstNit}
\end{figure}
To further investigate the deformations of \ce{Si3N4} waveguides upon high-fluence ion implantation, we implanted \ce{Si3N4} ridge waveguides without lateral cladding and characterized their geometries before and after implantation.
The \ce{Si3N4} ridge waveguides used in this test was fabricated by standard UV lithography and RIE etching of a 350-nm-thick LPCVD \ce{Si3N4} film on wet oxidized \ce{Si} wafer.
Irradiation energies and fluences ratio are kept the same as those used for implantation to lateral cladded waveguides.
The total fluence is $1\times10^{16}~\text{cm}^{-2}$.

We noted a significant change in the waveguide cross section upon ion implantation, by comparing the SEM images of waveguides cross sections before and after implantation (Fig.\ref{Fig:SIProc} (a) and (b)).
Clearly, the profiles of the waveguide sidewalls and the top surface after implantation are severely deformed.
The implanted waveguide cross sections appears to be wider and thinner compared to the masked waveguide, as shown in the top surface topography mapped by AFM (Fig.\ref{Fig:SubstNit} (d) and (2)).
Such deformations are more severe in comparison to those observed in implanted waveguides with lateral cladding (Fig.\ref{Fig:SIProc}(b),(c) and (d)), which highlights the importance and necessity of the mechanical support from the lateral \ce{SiO2} cladding.

\section{Selective masking for on-demand erbium implantation in complex \ce{Si3N4} photonic circuits.}

As commonly done in the current microelectronics industry for semiconductor doping, areas of the PICs that are not intended to be gain sections can be masked before the implantation in order to avoid modifying the property of passive waveguides.
Such selective implantation can sustain demanded properties such as ultra-low loss and engineered dispersion of passive devices, e.g., Kerr microring resonators, microring filters, couplers, and many others.
Erbium absorption loss in devices where the pump light is not deliverable can also be avoided.
This is a significant advantage over bulk doping during material growth reported in prior works using erbium-doped lithium niobate\cite{Zhou2021}.

\begin{figure}[htb]
	\centering
	\includegraphics[width=\textwidth]{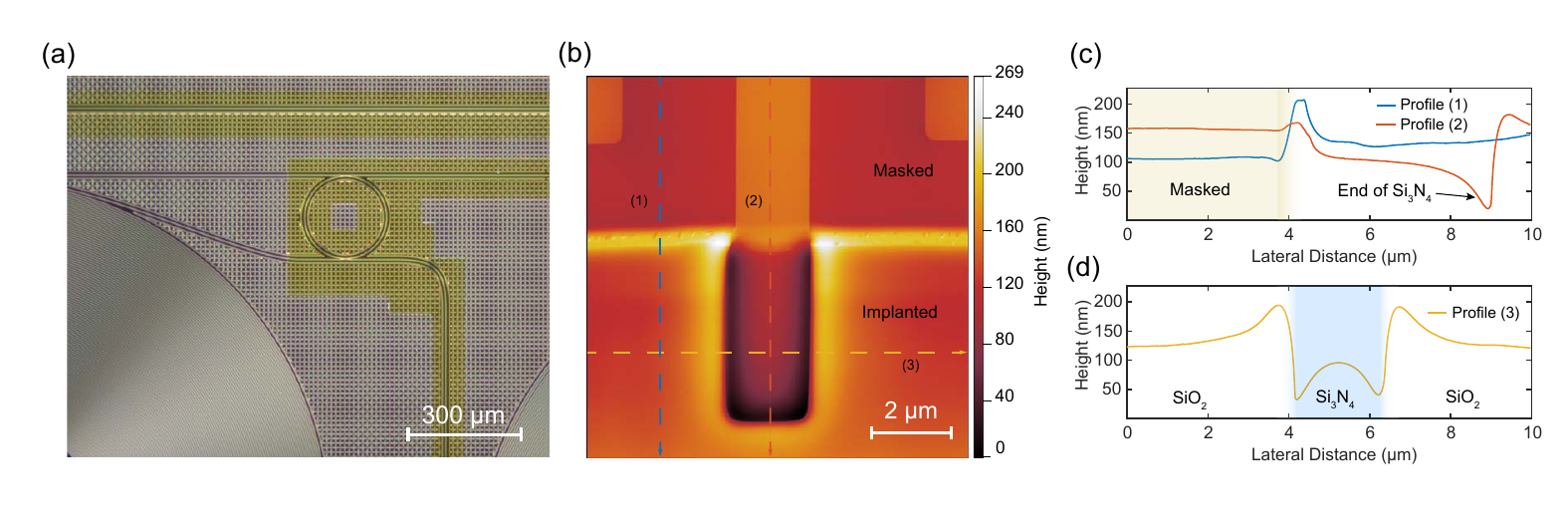}
	\caption{\textbf{Selectively erbium implanted \ce{Si3N4} photonic circuits.}
		\textbf{a}, Dark field optical microscope image of a typical selectively implanted photonic circuit.
		The region that contains a ring resonator and appears to be with a yellow tone is masked while the other area with spiral coil is implanted.
		The total implantation dose of this sample is around $1\times10^{16}~\text{cm}^{-2}$.
		\textbf{b}, Top surface topography of a \ce{Si3N4} section near the boundary between masked area and implanted area, mapped by AFM.
		\textbf{c}, Line profiles of (b) perpendicular to the boundary.
		\textbf{d}, Line profile of (b) parallel to the boundary, crossing the \SI{2}{\micro\metre} wide implanted \ce{Si3N4} region.
	}
	\label{maskedsample}
\end{figure}

We demonstrate the feasibility of selective masking for erbium ion implantation using a simple photoresist masking process.
In this process, we define the masked area by the standard UV direct-write lithography with \SI{3}{\micro\metre} thick AZ 15nXT photoresist before the implantation.
SRIM calculation indicates a \SI{2}{\micro\metre} photoresist layer is enough to stop all of the erbium ions at 2~MeV. 
After the erbium ion irradiation, we notice that the photoresist shows signs of strong cross-linking and appears to be brown in color.
We are able to completely remove the photoresist by ashing in high power oxygen plasma (15 min) and washing in \ce{HCl} solution ($37\%$, \SI{45}{\degreeCelsius}, 15 min).

Although extreme ion energy and ion dose are applied in the waveguide doping, i.e., the ion energy is comparable to implants used in the imaging sensor industry, and the dose is comparable to those for creating ohmic contacts in SiC processes\cite{Felch2013}. 
This simple masking process worked surprisingly well in our demonstration.
As shown in Fig.\ref{maskedsample}(a), after the photoresist is removed, a clear contrast between implanted and masked areas can be observed under an optical microscope, which provides evidence of material property modification by the ion irradiation. For the scope of future developments towards more complicated active \ce{Si3N4} photonic integrated circuits,  we also notice that the implanted areas can undergo some topography changes due to material volume expansion and stress accumulation.
As shown in Fig.\ref{maskedsample} (b), we observe minor local deformations of \ce{SiO2} at the boundary between the implanted area and masked area.
The implanted \ce{Si3N4} core appears to expand in volume, squeezing our adjacent lateral \ce{SiO2} cladding and creating small humps around the waveguide with a height variation on the order of 50~nm.
This observation agrees with the geometry change in cross section, as shown in Fig.\ref{Fig:SIProc} (c) and (d).

We simulated the fundamental TE optical mode field based on the SEM measured geometry shown in Fig.\ref{Fig:SIProc} (c) and (d) at 1550 nm. Assuming the transition between implanted waveguide cross section to unimplanted cross section is sudden, by computing the mode overlap, we estimated the fundamental mode to fundamental mode transmission loss at the boundary to be $<0.018$~dB and the return loss $<-45$~dB.
Although the reflection and scattering caused by the 'step-like' feature at the boundary are fairly small for many applications, we can apply methods such as introducing a grayscale mask or a tapering mask above the waveguide at the boundary to engineer a smooth transition and further reduce the reflection and scattering.

\section{Erbium ion distribution and overlap factor calculation. }

To ensure a maximum overlap of the erbium ion distribution and the optical intensity in the \ce{Si3N4} waveguides, we do three consecutive implantations of \ce{^{166}Er^+} at different energies and thus different ion ranges.
We optimize the implantation energies and fluences for the overlap according to the Stopping and Range of Ions in Matter (SRIM) calculations\cite{Ziegler2010}.
In the SRIM calculation, we used a density of 3.17 g/cm$^3$ for stoichiometric \ce{Si3N4}.
A Monte Carlo simulation with the Transport of Ions in Matter (TRIM) code is used later to obtain a more precise density distribution. 
The ion energies, TRIM simulated ion distribution parameters and the implantation fluences are listed in Table ~\ref{tbl:fluence}. 
We would like to note that the fluence received by some earlier samples is lower than expected by a factor of $0.42$ or even lower due to a sample mounting issue.
Rutherford backscattering spectroscopy (RBS) and optical absorption measured in ring resonators are used to calibrate the actual erbium concentration, which has been presented in the main manuscript.

\begin{table*}[htbp]
	\centering
	\def\arraystretch{1.3}
	\caption{\textbf{Parameters applied in the three consecutive erbium implantation into \ce{Si3N4} waveguides.}}
	\begin{tabular}{cccc}
		\toprule
		Energy (MeV)& Ion range (nm) & Straggling (nm)  & Ion fluence (ions/cm\textsuperscript{2}) \\
		\hline
		0.955                      & 197.4     & 40.0     & 2.34$\times$10\textsuperscript{15}   \\
		1.416                     & 284.9     & 55.3      & 3.17$\times$10\textsuperscript{15}      \\
		2.000                     & 398.1     & 74.1        & 4.5$\times$10\textsuperscript{15}      \\
		\hline
	\end{tabular}
	\label{tbl:fluence}
\end{table*}

Benefiting from the large doping depth (comparable to the waveguide thickness) and optimized vertical distributions, the ion implantation approach allows to achieve a large overlap factor between the erbium ions and optical modes.
Based on the parameters provided in Table \ref{tbl:fluence}, we estimate the overlap factor $\Gamma$ between erbium density distribution and the intensity of the fundamental TE mode of interest at 1550 nm to be $\sim$0.49, in a waveguide with a transversal dimension of \SI{2.1}{\micro\metre} $\times$ \SI{0.7}{\micro\metre}.
One can note that the actual overlap factor can be higher up to $\sim0.6$ for thinner waveguides, considering fabrication-induced thickness variation of samples.

Here we define the maximum Erbium concentration normalized $\Gamma$ as
\begin{equation}
	\label{eqn:gamma}
	\Gamma=\frac{\iint N(y) \epsilon(x,y) E^2(x,y) dxdy}{\max (N(y)) \iint \epsilon(x,y) E^2(x,y) dxdy}
\end{equation}
where $E(x,y)$ is the electric field of the fundamental TE mode, $\epsilon(x,y)$ is material relative permittivity and $N(y)$ is the depth-dependent ion concentration.

\section{Rutherford backscattering spectroscopy (RBS) measurement and analysis.}
\begin{figure*}[h!]
    
	\centering
	\includegraphics[width=\textwidth ]{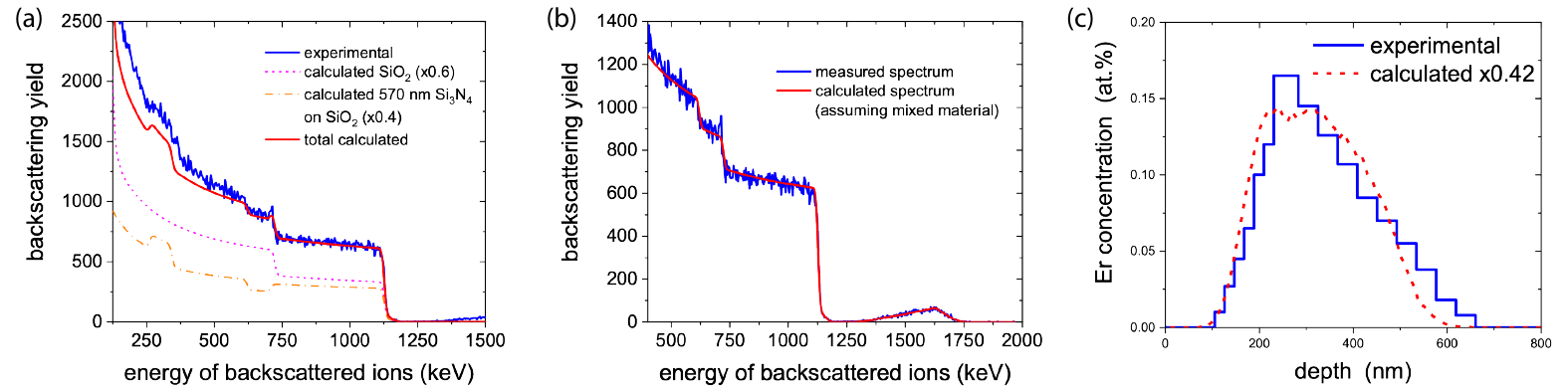}
	\caption{\textbf{Rutherford backscattering spectroscopy measurement and analysis}.  \textbf{a},  Part of the measured RBS spectra representing the elements of the matrix together with simulated RBS spectra. 
	\textbf{b}, Part of the RBS spectra highlighting the erbium signal together with a multi-layer simulated spectrum. 
	\textbf{c}, Extracted measured Er concentration profile (blue line) together with the respective SRIM simulation of the implantation profile.  
	}
	\label{Fig:SI_RBS}
\end{figure*}

The Er concentration profile of the implanted and annealed sample from the same wafer has been analyzed by Rutherford backscattering spectrometry (RBS) measurements using 2~MeV He ions. 
The ion beam diameter was about 1 mm, the backscattered ions were detected at $170^{\circ}$ in respect to the incident ion beam and the total ion charge was \SI{5}{\micro\coulomb} for collecting one spectrum. 
The measured spectrum (blue lines in Fig.\ref{Fig:SI_RBS}(a)) was calibrated by measuring additionally a thin surface layer sample containing several elements.    

As the sample consists of exposed areas of \ce{Si3N4} and \ce{SiO2}, firstly the relative coverage area of them had to be determined. 
This is done by fitting the simulated RBS spectrum (red line Fig.\ref{Fig:SI_RBS}(a)) to the respective Si, N and O signals appearing at about 1125, 730 and 625~keV, respectively, which results to a coverage fraction of \ce{SiO2}:\ce{Si3N4} = 0.4:0.6, as shown in Fig.\ref{Fig:SI_RBS}(a). 
Next, the Er signal located between 1350 and 1750~keV and clearly visible in Fig.\ref{Fig:SI_RBS}(b) was analyzed in detail. 
A very good fit of simulated and measured RBS spectra was obtained by assuming layers with different Er concentrations of thicknesses ranging from 20-40~nm, which represents the depth resolution of the performed RBS measurements.
As RBS measures the density profile in thin film units, the depth scale in Fig.\ref{Fig:SI_RBS}(c) of the measured Er concentration is converted to physical depth assuming the density of \ce{Si3N4} to be $9.65\times 10^{22}~\text{atoms/cm}^3$ (3.17 g/cm$^3$). 
We obtain a good agreement of the shape of the measured profile with the one calculated by SRIM, also shown in main manuscript, but the absolute erbium concentration measured is lower by a factor of 0.42 than expected from the nominal implantation fluences.  

\section{Loss characterization of erbium-implanted \ce{Er\dop Si3N4} waveguides.}
The wavelength-dependent waveguide losses of the \ce{Er\dop Si3N4} circuits need to be carefully characterized, in order to infer the waveguide background loss, erbium absorption and the net gain coefficient.
Here, we use two methods, i.e. the microring resonance linewidth characterization and the optical frequency-domain reflectometry (OFDR), to calibrate the waveguide losses.

\begin{figure*}[h!]
    
	\centering
	\includegraphics[width=\textwidth ]{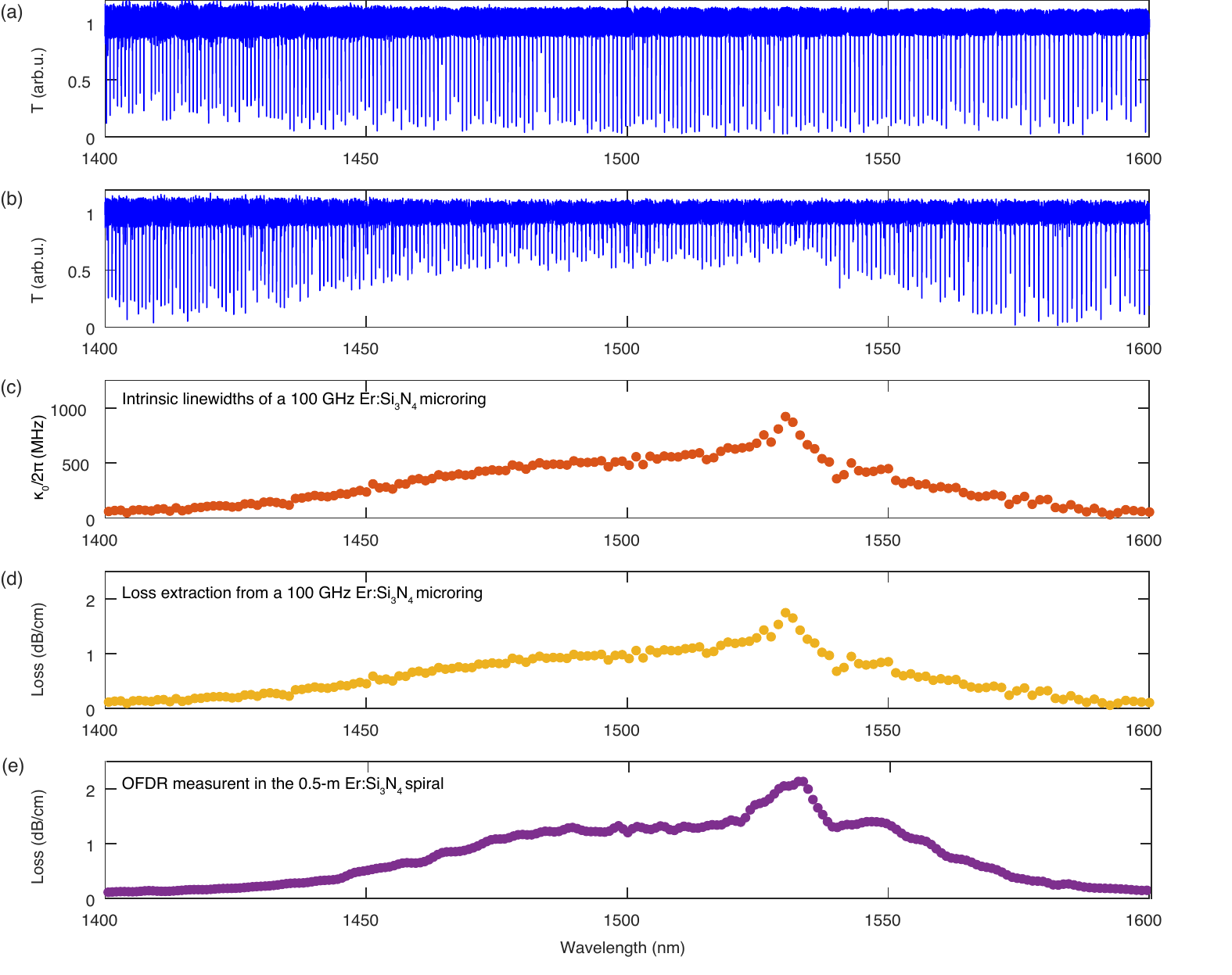}
	\caption{\textbf{\ce{Er\dop Si3N4} waveguide loss characterization}.  \textbf{a},  Normalized transmission of a 100 GHz \ce{Si3N4} microring resonator before erbium implantation. \textbf{b}, Normalized transmission of a 100 GHz FSR \ce{Er\dop Si3N4} microring resonator after \SI{1000}{\celsius} annealing for erbium activation and loss recovery. \textbf{c}, Intrinsic linewidths ($\kappa_0/2\pi$) of the annealed \ce{Er\dop Si3N4} microring resonator. \textbf{d}, Waveguide losses extracted from the intrinsic linewidths.
\textbf{e}, Waveguide losses obtained from the OFDR measurement in the 0.5-m-long \ce{Er\dop Si3N4} waveguide spiral.
	}
	\label{Fig:SI_kappa2loss}
\end{figure*}

We first characterize the resonance linewidth of a 100 GHz \ce{Er\dop Si3N4} microring resonator side coupled with a single bus waveguide.  
Fig.\ref{Fig:SI_kappa2loss}(a) and (b) show the optical transmission spectra of the microring before the erbium ion implantation and after the implantation and annealing, respectively, using a home-built sub-MHz-resolution optical network analyzer\cite{Liu2016,Riemensberger2021}. 
Variations in the resonance dip depth are observed, which are associated with changes in intrinsic linewidth $\Delta \nu_0 = \kappa_0/2\pi$.  
The intrinsic dissipation rate $\kappa_0$ is inferred from fitting of the measured resonance to a Lorentzian lineshape, as shown in Fig. \ref{Fig:SI_kappa2loss}(c).  
The optical loss $\alpha$~(in 1/m) is converted from the intrinsic dissipation rate via the expression given by
\begin{equation}
\alpha = \frac{n_\mathrm{eff}\kappa_0}{c},
\label{eqn:loss}
\end{equation}
where $n_{eff}$ is the effective refractive index and $c$ is the light speed in vacuum. The calculated waveguide loss is shown in Fig. \ref{Fig:SI_kappa2loss}(d), indicating a peak value of $\sim$1.75~dB/cm at 1535~nm

Secondly, we characterize the propagation loss in a 0.5-m-long \ce{Er\dop Si3N4} waveguide via OFDR measurements. 
The OFDR traces record the optical back reflection along the waveguide length, which can be converted to propagation loss by linear fitting of the distance-dependent optical backreflectance. 
By fitting the loss in various segmented spectral windows, the wavelength-dependent loss is obtained, as shown in Fig. \ref{Fig:SI_kappa2loss}(e). 
The characterized loss from the OFDR measurement in general agrees with the loss inferred from the intrinsic resonance linewidth measurement, while exhibiting a slightly higher peak value of 2.14~dB/cm at 1535~nm. This difference is attributed to the possible reduced erbium absorption caused by the cavity enhanced optical intensity for the case of resonance linewidth measurements. In gain calculations and simulations, we use the losses inferred from the OFDR measurements.

\section{Photoluminescence lifetime measurement.}
The photoluminescence (PL) lifetime $\tau$ of the transition from the excited state $^4I_{13/2}$ to the ground state $^4I_{15/2}$ is measured from a 0.45 cm-long \ce{Er\dop Si3N4} waveguide, in order to avoid the possible erbium re-absorption in long waveguides.
A 980 nm pump laser diode is gated by a periodic square wave with a duty cycle of 50$\%$ and a duration of 50 ms.
The coupled pump power is limited to $<$1~mW level to minimize the effect from cooperative upconversion processes\cite{VanDenHoven1996a}.
When the modulated pump is switched off, the spontaneous emission decays at a rate inversely proportional to the PL lifetime $\tau$.
An increased lifetime $\tau$ of 3.4~ms for the decay to its $1/e$ of the maximum is obtained from exponential curve fitting, compared to 1.5~ms before annealing.

\section{Gain coefficient analysis for different pump wavelengths and waveguide cross sections. }

\begin{figure*}[htbp]
	\centering
	\includegraphics[width=\textwidth ]{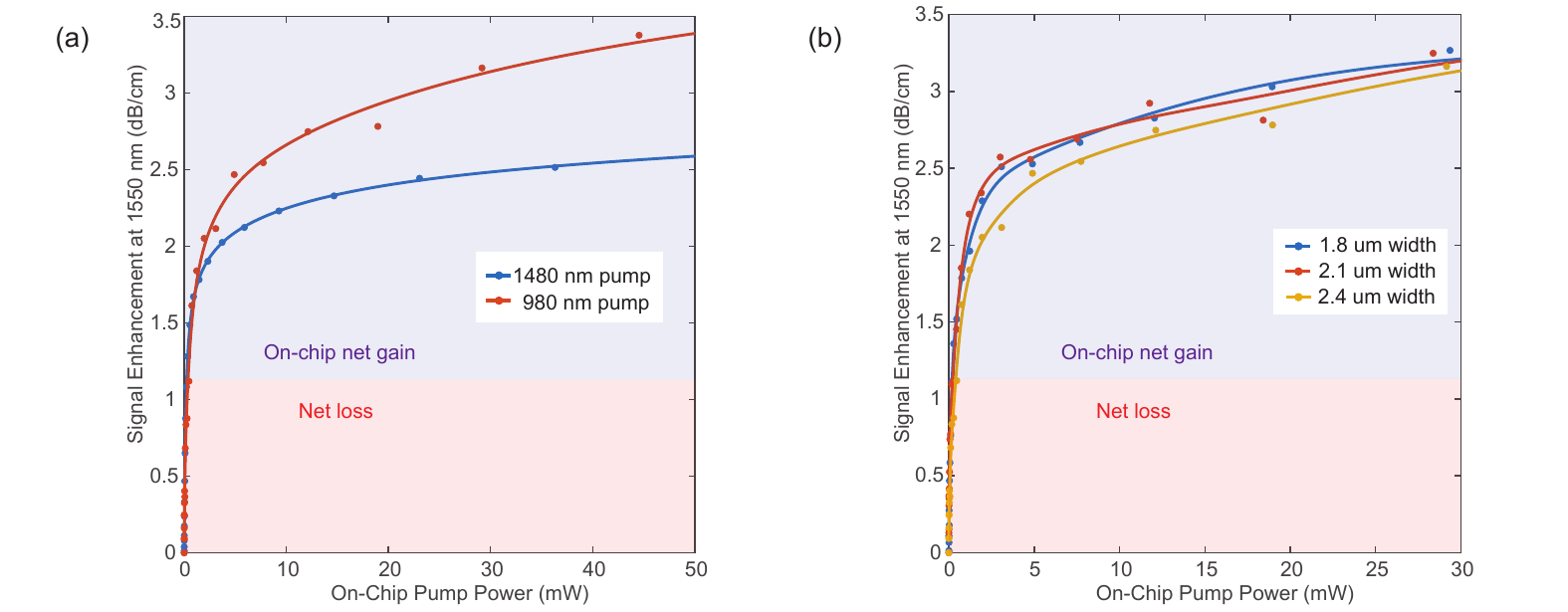}
	\caption{\textbf{Signal gain dependence on the pump wavelength and the waveguide width}.  \textbf{a},  Measured signal enhancement per centimeter \ce{Er\dop Si3N4} waveguide upon pumping at 980 nm and 1480 nm. \textbf{b}, Measured signal enhancement per centimeter for \SI{1.8}{\micro\metre}, \SI{2.1}{\micro\metre}, and \SI{2.4}{\micro\metre} wide waveguide upon 1480 nm pumping. Solid dots are measured and lines are guides for the eye created by smoothing spline. The samples are doped with $\sim 1.35\times  10^{20}~\text{cm}^{-3}$ peak concentration.
	}
	\label{Fig:SIgain}
\end{figure*}

Firstly, we investigate the gain coefficients in a short (0.43 -- 0.46 cm) \ce{Er\dop Si3N4} waveguide at different pumping wavelength, as shown in Fig. \ref{Fig:SIgain} (a).
A maximum signal enhancement of 3.4~dB/cm at 1550 nm is obtained for 980~nm pumping, which is higher than the peak value of 2.5~dB/cm upon 1480~nm pumping, due to the more complete population inversion upon 980~nm pumping.
A net gain coefficient of 2.3~dB/cm for 980~nm pumping is obtained by subtracting the total losses from erbium absorption and waveguide background loss, while a net gain coefficient of 1.4~dB/cm for 1480~nm pumping.
The theoretic gain coefficient is given by $g=\sigma_\mathrm{e}\Gamma N_0$ (1.7~dB/cm) \cite{Giles1991}. 
For experiments presented in the manuscript, 1480~nm pumps are deployed due to the lower coupling loss and the cut-off wavelength (1450~nm) of the UHNA fibers for chip coupling.

Secondly, Fig. \ref{Fig:SIgain} (b) compares the optical net gain of waveguides doped with the same erbium ion fluence, but different widths. The saturated gain of the three waveguides with \SI{1.8}{\micro\metre}, \SI{2.1}{\micro\metre}, and \SI{2.4}{\micro\metre} width is the same within the measurement error. 
From the gain-pump power relation, we find that the \SI{1.8}{\micro\metre} and \SI{2.1}{\micro\metre} waveguides reach saturation at the same rate while the \SI{2.4}{\micro\metre} waveguide has a slightly higher pump saturation power due to the relatively larger effective mode area. 
The gain performance can be further enhanced by adopting increased waveguide cross section dimensions to achieve a larger mode area for higher output saturation power. 
This result indicates that \ce{Er\dop Si3N4} waveguides with \SI{1.8}{\micro\metre} to \SI{2.4}{\micro\metre} widths have similar amplification performance, which permits flexible design of waveguide geometry for optical dispersion engineering for applications such as mode-locked laser and active cavity solitons.

In experiments, we measure the power of 1550~nm signal before and after passing the short waveguides pumped with varied optical power.
We derived the signal enhancement which is the difference of the maximum and the minimum output signal power.
The net optical gain is then computed by subtracting the characterized erbium absorption loss at 1550~nm from the measured signal enhancement. 
We set the on-chip input signal power at $-31$ dBm to avoid gain saturation and ensured the output signal level is higher than the total power of amplifier spontaneous emission (ASE) power by at least 5 dB. 
The ASE level is calibrated without the signal input, which is then subtracted from the measured total output signal power.
Wavelength division multiplexers (Thorlabs WD1450A) and free-space thin-film optical filters (Thorlabs FB1550-40 and FEL1500) are used to suppress residual pump light by $-60$ dBc in output signal power measurements.

\section{Numerical modeling of optical amplification in \ce{Er\dop Si3N4} waveguides.}

\begin{figure}[!htb]
	\centering
	\includegraphics[width=0.8\textwidth]{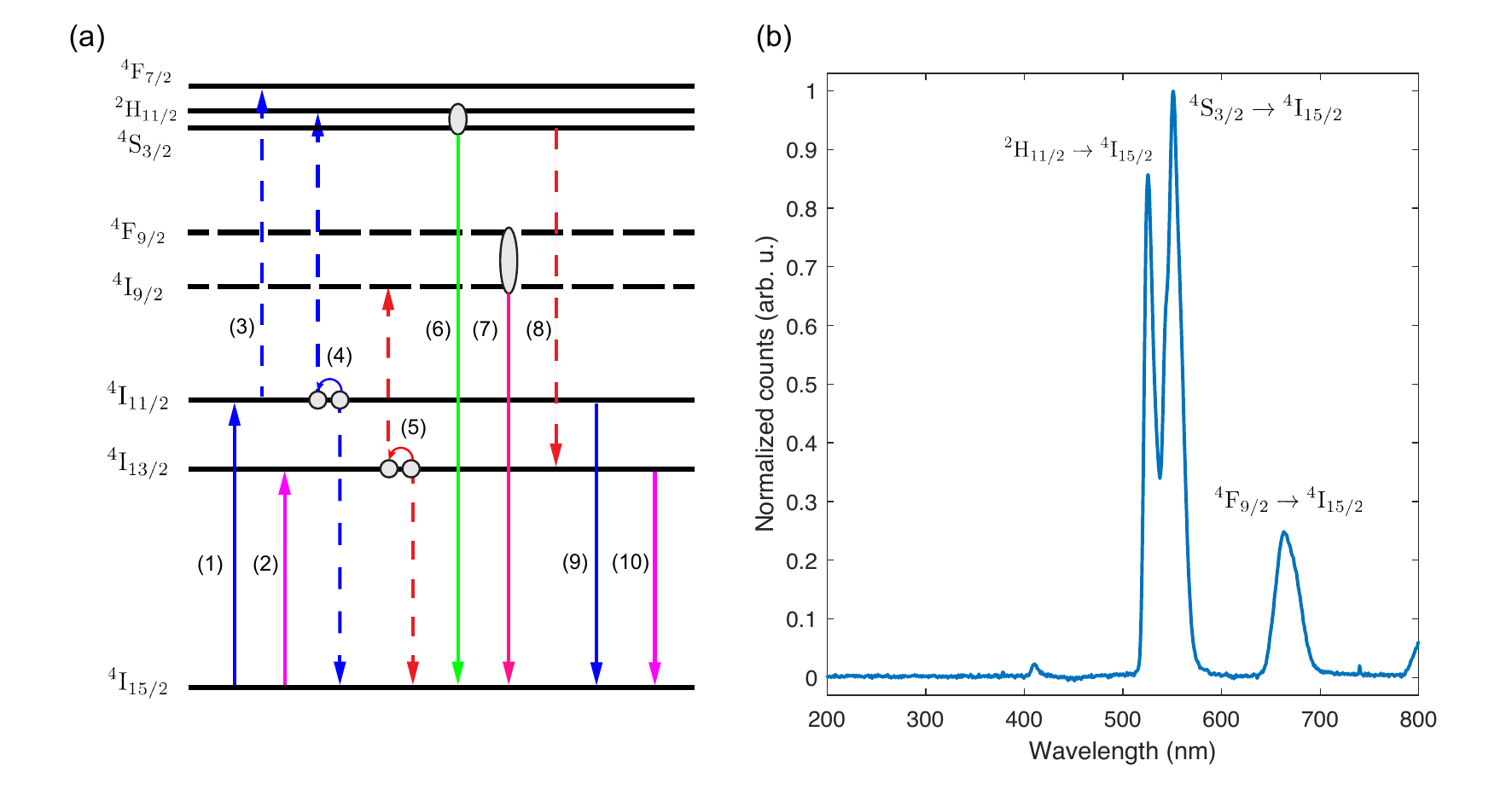}
	\caption{\textbf{Energy levels and up-conversion transitions in \ce{Er\dop Si3N4}}. \textbf{a}, Scheme of energy level of \ce{Er^{3+}} ions in solids and important transitions in between.
		Non-radiative relaxation are omitted for clearness.
		Transitions labeled: (1) 980 nm pump, (2) 1480 nm pump, (3) excited state absorption (ESA) of 980 nm pump, (4) cooperative up-conversion from  ${}^4\text{I}_{11/2}$, (5) cooperative up-conversion from ${}^4\text{I}_{13/2}$, (6) Green emission at 520 nm/545 nm, (7) Red/NIR emission at 665 nm/800 nm, (8) 845 nm emission, (9) 980 nm emission, (10) 1530 nm emission used for C-band optical amplification. \cite{Arahira1992,Kir2013}
		\textbf{b}, Measured optical spectrum (intensity uncalibrated) of multi-step up-converted luminescence emitted by the \ce{Er\dop Si3N4} spiral waveguide under intense 1480 nm pump. Featuring emission (6) and partially (7). The weak emission at 410 nm may be attributed to a three step up-conversion to a higher level ${}^2\text{H}_{9/2}$\cite{Arahira1992, Kik2003}
		}
	\label{Fig:erlevels}
\end{figure}

We model the optical gain by treating the erbium as a three-level system that generally well describes the populations in the first three lower-lying levels (${}^4\text{I}_{15/2}$,${}^4\text{I}_{13/2}$ and ${}^4\text{I}_{11/2}$), as shown in Fig. \ref{Fig:erlevels}(a).
The rate equations that govern the population dynamics including excited state absorption (ESA), first- and second-order cooperative upconversions (CUC) upon pumping with 1480~nm are given by\cite{Simpson1999_book,Kik2003},

\begin{equation}
	\begin{aligned}
		\frac{dN_3}{dt} &= -(A_{32} + C_{37}N_3) N_3 + (R_{24} + C_{24}N_2)N_2,\\
		\frac{dN_2}{dt} &= -(A_{21}  + R_{21} + R_{24} + 2C_{24}N_2)N_2 + R_{12}N_1  + A_{32}N_3,\\
		\frac{dN_1}{dt} &=  - R_{12}N_1 + (A_{21}  + R_{21}+ C_{24}N_2)N_2 + C_{37}N_3^2,
	\end{aligned}
	\label{eq:threelevel}
\end{equation}
where $N_{1,2,3}$ indicates the populations at the first three levels ${}^4\text{I}_{15/2}$ , ${}^4\text{I}_{13/2}$ and ${}^4\text{I}_{11/2}$ respectively.
$A_{ij} = 1/\tau_{ij}$ is the decay rate from level $i$ to level $j$, given by the inverse of the life time $\tau_{ij}$.
$R_{ij}=\sigma_{ij}\phi$ is the transition rate of optical beams (signal, pump and spontaneous emission) from state $i$ to state $j$, where $\phi = \frac{I}{h\nu_k}$ is the photon flux, $\sigma_{ij}$ is the transition cross section, $I$ is the optical intensity and $v$ is the photon frequency.
$R_{24}$ presents the transition rate of ESA from the ${}^4\text{I}_{11/2}$ level, while $C_{24}$ and $C_{37}$ are responsible for the first- and second-order cooperative upconversions, respectively.
These terms are included in the numerical calculations, although it has been demonstrated that $C_{37}$ could be small for ion implanted substrate with respect to co-sputtered doped film, due to the relatively homogeneous erbium distribution\cite{Kik2003}. $N = N_1 + N_2 + N_3$ is the total ion concentration.

In the waveguide, the spatial distributions of the signal, pump and the spontaneous amplifier noises can be described by a set of propagation Eq given by

\begin{equation}
	\begin{aligned}
		\frac{dP_{\mathrm s}(z)}{dz}&=P_{\mathrm s}(z)\Gamma_{\mathrm s}\left[\sigma_{21\mathrm s}N_2(z)-\sigma_{12\mathrm s}N_1(z)\right] -P_\mathrm{s}(z)\alpha_{0\mathrm s},\\
		\frac{dP_{\mathrm p}(z)}{dz}&=P_{\mathrm p}(z)\Gamma_{\mathrm p}\left[\sigma_{21\mathrm p}N_2(z)-\sigma_{12\mathrm p}N_1(z)\right] -P_\mathrm{p}(z)\alpha_{0\mathrm p},\\
		\frac{dP_\mathrm{ASE,f}(z)}{dz}&=P_\mathrm{ASE,f}(z)\Gamma_\mathrm{ASE}\left[\sigma_{21,\mathrm{ASE}}N_2(z)-\sigma_{12,\mathrm{ASE}}N_1(z)\right]
                                          +P_\mathrm{ASE}^0\sigma_{21,\mathrm{ASE}}N_2(z) -P_\mathrm{ASE,f}(z)\alpha_\mathrm{0,ASE},\\
      	\frac{dP_\mathrm{ASE,b}(z)}{dz}&=-P_\mathrm{ASE,b}(z)\Gamma_\mathrm{ASE}\left[\sigma_{21,\mathrm{ASE}}N_2(z)-\sigma_{12,\mathrm{ASE}}N_1(z)\right]
      									 -P_\mathrm{ASE}^0\sigma_{21,\mathrm{ASE}}N_2(z)  +P_\mathrm{ASE,b}(z)\alpha_\mathrm{0,ASE},\\
	\end{aligned}
	\label{eqn:pp}
\end{equation}
where $\Gamma$ is the mode-erbium overlap factor, $\alpha_\mathrm{0s}$ and $\alpha_\mathrm{0p}$ are the waveguide background losses for the signal and pump, respectively.  $P_\mathrm{ASE}^0 = 2Bh\nu_\mathrm{ASE}$ is the added local spontaneous emission noise power within a bandwidth of $B$ ($\sim$5~THz in simulations) for two polarizations.  
In simulations, we include the amplifier spontaneous emission noises propagating in both directions.
	For the cases with counter-propagating optical beams such as in backward pumping and and bidirectional pumping configurations, Equations \ref{eqn:pp} need to be expanded to include both forward and backward propagating pumps, and can be solved as an ordinal differential equation boundary value problem by algorithms such as the relaxation algorithm.

\begin{figure}[!h]
	\centering
	\includegraphics[width = \textwidth]{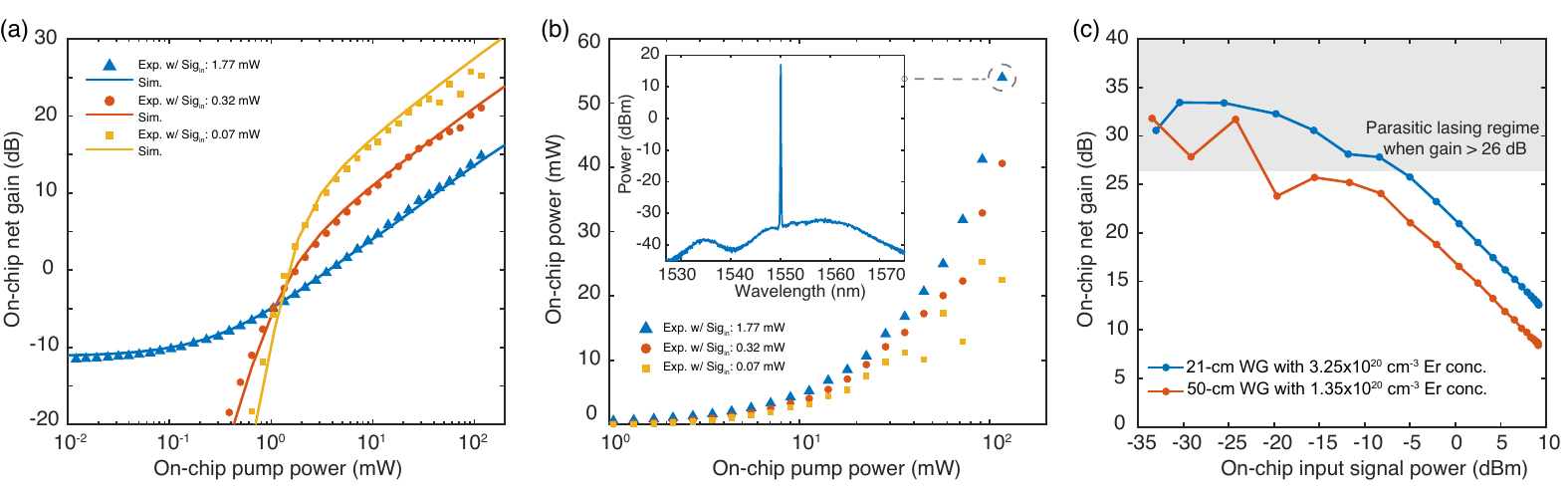}
	\caption{\textbf{Measured and numerical fitted optical amplification in \ce{Er\dop Si3N4} spiral waveguides. } 
	\textbf{a}, Measured (scatters) and simulated (solid curves) on-chip net gain for signals at 1550 nm in the 50-cm-long \ce{Si3N4} waveguide doped with $1.35\times 10^{20}~\text{cm}^{-3}$ erbium concentration. 
\textbf{b}, The corresponding on-chip output powers at 1550 nm.  The inset shows the calibrated optical spectrum of the 55~mW signal output at a pump power of $\sim$120~mW at 1480~nm.  
\textbf{c}, On-chip net gain as a function of input signal power at 1550~nm for two \ce{Er\dop Si3N4} waveguide amplifiers with different erbium concentrations.
The shaded area indicates the regime where parasitic lasing takes place when the net gain exceeds $\sim$26~dB.
}
	\label{Fig:sim_gain_50cm}
\end{figure}

Figure \ref{Fig:sim_gain_50cm}(a) shows the numerical fitting to the measured optical gain using simulations based on Equations \ref{eq:threelevel} and Equations \ref{eqn:pp},  in order to estimate the contributions from ESA and CUC processes. 
The corresponding output powers are shown in Fig. \ref{Fig:sim_gain_50cm}(b). 
The experimental results presented in the main manuscript are reproduced using the parameters listed in Table \ref{Tab:Sim_pars}. 
The simulations suggest that the observed optical gain is not obviously limited by the second order cooperative upconversion $C_{37}$, even in the regime with $>$ 120~mW on-chip pump power due to the small $C_{37}$ from the numerical fitting of measured results.  
Despite the weak $C_{37}$, we observed the green luminescence induced by the transition from higher-lying levels upon intense pumping, as shown in Fig. ~\ref{Fig:erlevels}(b).  Figure \ref{Fig:sim_gain_50cm}(c) shows the net gain as a function the input signal power $P_\text{sat,s} $ at 1550~nm, from which the input saturation power is estimated to be around -15~dBm (where small-signal gain drops by 3 dB),  approximately matched with the theoretical value of -14~dBm given by $P_\text{sat,s} = \frac{hv_\mathrm{s}}{(\sigma_\mathrm{21s}+\sigma_\mathrm{12s})\tau_{21}A_\text{eff}}$ where the $A_\text{eff}=1.2 \mu\text{m}^2$ is the effective area of the fundamental TE mode. The shaded area indicates the regime where parasitic lasing kicks in when the net gain exceeds $26$~dB, which leads to fluctuations in measured net gain and causes uncertainty in the estimation of saturation input power. The vertical offset between the net gains from two samples is mainly caused by the difference in coupled pump powers.

\begin{figure}[!h]
	\centering
	\includegraphics[width=0.75\textwidth]{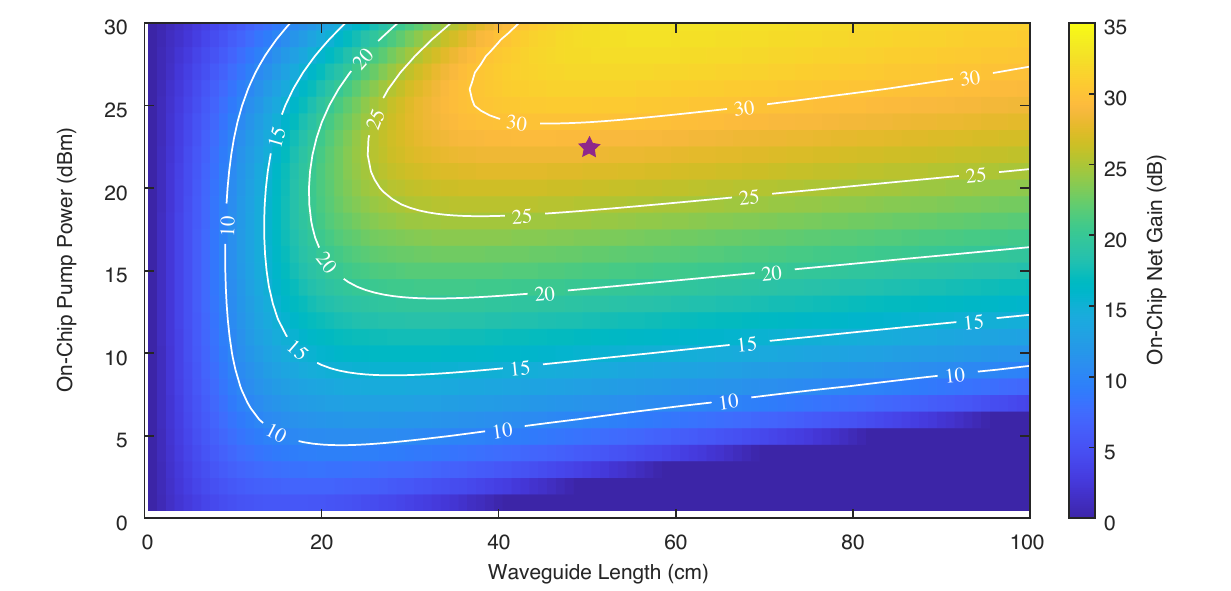}
	\caption{\textbf{Calculation of on-chip net gain under various pump powers at 1480~nm and different waveguide lengths for $-10$~dBm on-chip input signal at 1550~nm. } Based on experimentally extracted cross sections and fitted parameters, the achievable on-chip net gain is calculated in an \ce{Er\dop Si3N4} with the length varied from 5~mm to 1~m with an on-chip forward pump power of up to 1~W. The star marker indicates the regime around which the \ce{Er\dop Si3N4} waveguide amplifier can approach in the presented demonstrations. Further increase in optical gain can be achieved by deploying higher pump powers or reducing the coupling loss.}
	\label{Fig:sim_2D}
\end{figure}

Figure \ref{Fig:sim_2D} presents the simulated on-chip net gain as functions of the pump power and the waveguide lenghth.  
An optimized waveguide length of 0.5-m can be found for an \ce{Er\dop Si3N4} amplifier for the demonstrated erbium concentration of $\sim1\times10^{20}\mathrm{cm^{-3}}$, capable of delivering an optical net gain exceeding 30~dB and an on-chip output power of 20~dBm (100~mW) with an on-chip power pump of around 23~dBm. 
The further increase of the waveguide length would not significantly contribute to a higher net gain, as the signal re-absorption by erbium can take place in the presence of the pump depletion and attenuation in longer waveguides.
In experiments, due to the power limit of the laser diode and coupling loss, the \ce{Er\dop Si3N4} waveguide amplifier can operate around the regime indicated by the marker shown in Fig. \ref{Fig:sim_2D}, for achieving 26~dB gain. 
Indeed, it is feasible to further increase the gain using a higher on-chip pump power exceeding 23~dBm, however, it would be limited to $\sim$30~dB by the excited state absorption (determined by $\sigma_{24}$) according to the simulations. 
In contrast, no significant contributions from cooperative upconversion processes determined by $C_{24}$ and $C_{37}$ are observed in simulations at pump levels of $>$200~mW. 
The fitted coefficients $C_{24}$ and $C_{37}$ of the samples based ion implantation are two orders of magnitude lower than what have been reported in sputtered erbium-doped \ce{Al2O3}\cite{Kik2003,VanDenHoven1996a,Ronn2019}.

\begin{table}
\caption{\textbf{Parameters utilized in simulations of on-chip optical net gain in a 0.5-m \ce{Si3N4} waveguides for -10~dBm input at 1550~nm.}}
\def\arraystretch{1.2}
\resizebox{\linewidth}{!}{%
\centering
\begin{tabular}{ccccccccccc} 
\toprule
Parameter & $N_0 (\mathrm{cm}^{-3})$    & $\tau_{21}$ (ms)& $\tau_{32}$ (\SI{}{\micro\second}) & $\sigma_{21\mathrm{p}} (\mathrm{cm}^{-2})$     & $\sigma_\mathrm{12p} (\mathrm{cm}^{-2})$     & $\sigma_\mathrm{21s} (\mathrm{cm}^{-2})$    &$\sigma_\mathrm{12s} (\mathrm{cm}^{-2})$     & $\sigma_{24} (\mathrm{cm}^{-2})$        & $C_{24} (\mathrm{cm^{3}s^{-1}})$  & $C_{37} (\mathrm{cm^{3}s^{-1}})$  \\ 
\hline
Value     & $1.35\times10^{20}$ & 3.4 & 20 & $1.07\times10^{-21}$ & $4.48\times10^{-21}$ &$6.76\times10^{-21}$ & $4.03\times10^{-21}$  & $1.0\times10^{-22}$  &$3.0\times10^{-18}$ & $1.0\times10^{-18}$             \\
\hline
\end{tabular}
}
\label{Tab:Sim_pars}
\end{table}

\section{Experimental setup for gain and output power measurements.}
The integrated waveguide amplifier is excited by pump lights provided by multi-longitudinal-modal laser diodes centered around 1480 nm (400~mW power at the fiber pigtail output), which are injected to the photonic chip via coupled fibers in both forward and backward directions. 
Here, we use 1480~nm pumping due to the lower fiber-to-chip coupling loss, despite a higher gain coefficient with 980~nm pumping.
The probe signal emitted from a frequency-tunable continuous-wave (CW) laser is coupled to the chip, after combined with the forward optical pump via a fiber-based coupler (wavelength-division multiplexer for 1480/1550~nm). 
The transmitted pump is filtered out by a high-extinction free-space low-pass filter, to ensure accurate gain characterization.  The measured signal and pump powers are carefully calibrated with characterized insertions losses caused by fiber links, optical couplers and optical filters. 
The optical spectra are calibrated with the optical powers measured by power meters.
We extract the net gain by directly comparing the output and input signal powers after calibrations, instead of using subtraction between signal enhancement and total loss which might lead to overestimation of the net gain.

\section{Fresnel reflection at chip facets and induced parasitic lasing. }
\begin{figure*}[h!]
	\centering
	\includegraphics[width=\textwidth ]{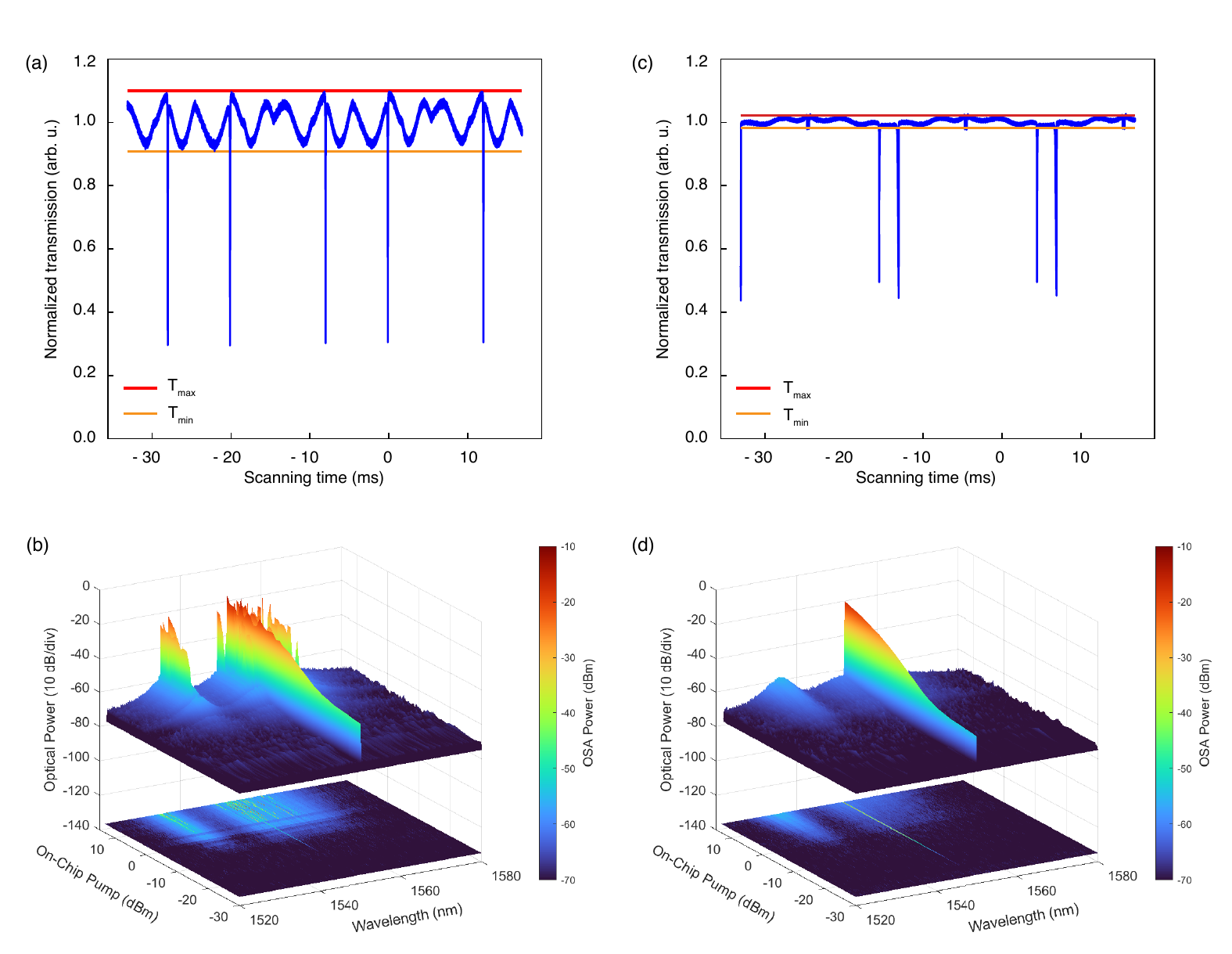}
	\caption{\textbf{Reduction of Fresnel reflection and parasitic lasing}.  \textbf{a},  Normalized transmission of a passive \ce{Si3N4} microring resonator side coupled with a bus waveguide with a length of 0.5~cm.  A pair of lensed fibers are used for light coupling to the chip, with air gaps.  The calculated Fresnel reflection is 4.77$\%$.
\textbf{b}, Measured optical spectra and the projection of the CW signal amplified by a 0.5-m-long \ce{Er\dop Si3N4} waveguide interfaced with lensed fibers. Parasitic lasing is clearly observed with $>$10 mW on-chip pump power.
 \textbf{c},  Normalized transmission of a passive \ce{Si3N4} microring resonator side coupled with a bus waveguide with a length of 0.5~cm.  A pair of UHNA fibers have close contact with the waveguide facets for light coupling, and the gaps are filled with refractive index matching gel. The calculated Fresnel reflection is 0.5$\%$.
\textbf{d}, Measured optical spectra and the projection of the CW signal amplified by a 0.5-cm-long \ce{Er\dop Si3N4} waveguide coupled to UHNA fibers with refractive index matching gel applied. Parasitic lasing is suppresed. The input signal power is kept constant at $\sim$ -10~dBm for both cases.
	}
	\label{Fig:SI_parasitic_lasing}
\end{figure*}

The Fresnel reflection of the waveguide input and output at the chip facets can form an optical Fabry-P\'erot (FP) cavity that can lead to resonance enhancement at resonance frequencies.
Fig. \ref{Fig:SI_parasitic_lasing}(a) show the transmission of a passive 100~GHz FSR \ce{Si3N4} microring resonator coupled with a 0.5-mm-long bus waveguide when using lensed fibers for light coupling in and out of the chip.
The laser frequency is scanned over 30~GHz with a period of 20~ms.
The baseline variation stemmed from the FP cavity fringes.
Assuming the same reflectance for both facets and negligible waveguide loss, the intensity reflection coefficient $R$ can be inferred from the contrast $K$ of the fringes via the expression\cite{Feuchter1994}
\begin{equation}
R = \frac{1-\sqrt{1-K^2}}{K}.
\label{eqn:R_fringes}
\end{equation}

The contrast is given by $K=\frac{I_\mathrm{max}-I_\mathrm{min}}{I_\mathrm{max}+I_\mathrm{min}}$ where $I_\mathrm{max}$ and $I_\mathrm{min}$ are the fringe maxima and minima, respectively.
This gives a calculated intensity reflection coefficient $R=4.77\%$, corresponding to a -13~dB return loss at each facet.
When the on-chip gain is sufficient to overcome the waveguide loss and the return loss, parasitic lasing of the F-P cavity can take place in the on-chip waveguide amplifier at wavelengths other than the input signal wavelength.
This effect will limit the achievable optical gain. 
In a 0.5-m-long \ce{Er\dop Si3N4} waveguide amplifier,  the parasitic lasing will limit the single-pass on-chip net gain to around 16~dB, when considering a 3~dB background loss. Fig.\ref{Fig:SI_parasitic_lasing}(b) shows an example of parasitic lasing for -10~dBm signal input at on-chip pump powers of $>$10~dBm.
With the unstable parasitic lasing, the amplified signal power exhibits fluctuations and does not show a clear increase when the pump power ramps up.

In contrast, the Fresnel reflection can be suppressed by reducing the refractive index mismatch caused by the air gap between the chip facets and coupled fibers.  
Here, we use a pair of ultra-high numerical aperture (UHNA) optical fibers that are butt-coupled against the input and output chip facets, respectively. 
The UHNA fibers are perpendicularly cleaved to provide flat end facets in contact with the \ce{Er\dop Si3N4} chip facets.  
The UHNA fiber (UHNA7, Coherent Inc.) has a small mode field diameter of 3.2~$\mu$m that approaches the optical mode size in the inversely tapered \ce{Si3N4} waveguides. 
We apply index matching gel (G608N3, Thorlabs Inc.) to the air gap between the UHNA fibers and the chip facets to reduce the Fresnel reflection by one order of magnitude from 4.77$\%$ to 0.5$\%$,  corresponding to a reflection loss of -23~dB, as shown in Fig. \ref{Fig:SI_parasitic_lasing}(c). 
The reduced reflection raise the threshold for parasitic lasing, which in turn allows for on-chip net gain up to 26~dB.  
In this case, no obvious parasitic lasing is observed in the same \ce{Er\dop Si3N4} waveguide with the same input signal, as shown in Fig. \ref{Fig:SI_parasitic_lasing}(d).  
The parasitic lasing can still takes place when the net gain exceeds 26~dB for either smaller input signals or increased pump powers, since the waveguide can provide sufficient high single-pass gain to surpass the FP cavity losses at one end, i.e. -23 dB for inward Fresnel reflection and -3 dB for waveguide background loss.
Further reduction of facet reflection by angled waveguide tapers and facets will facilitate a higher achievable net gain. 
In experiments,  fiber-to-chip coupling losses are 2.9~dB and 3.3~dB per side for 1550~nm and 1480~nm, respectively.

\section{Noise figure of the \ce{Er\dop Si3N4} amplifier.}
The noise figure of the \ce{Er\dop Si3N4} amplifier is measured using the commonly used optical source subtraction method\cite{Derickson1998} that can remove the source spontaneous emission (SSE) noise from the total noise emitted by the erbium amplifier.
In the measurement, the spectra of the input laser signal and the amplified signal recorded by an optical spectrum analyzer (OSA) are calibrated with respect to their powers measured by a power meter. 
The optical insertion loss between the amplifier output and the OSA is measured and then compensated in calculations. 
The optical gain $G$ at the optical frequency $v$ and noise powers at the input $P_\text{SSE}$ and the output $P_\text{out}$ within an optical bandwidth $B_0$ are measured from the calibrated optical spectra, in order to derive the noise figure using the expression given by
\begin{equation}
\mathrm{NF} = \frac{P_\text{out} - GP_\text{SSE}}{GhvB_0} + \frac{1}{G}.
\label{eqn_SI: NF}
\end{equation}

\begin{figure*}[h]
	\centering
	\includegraphics[width=0.8\textwidth ]{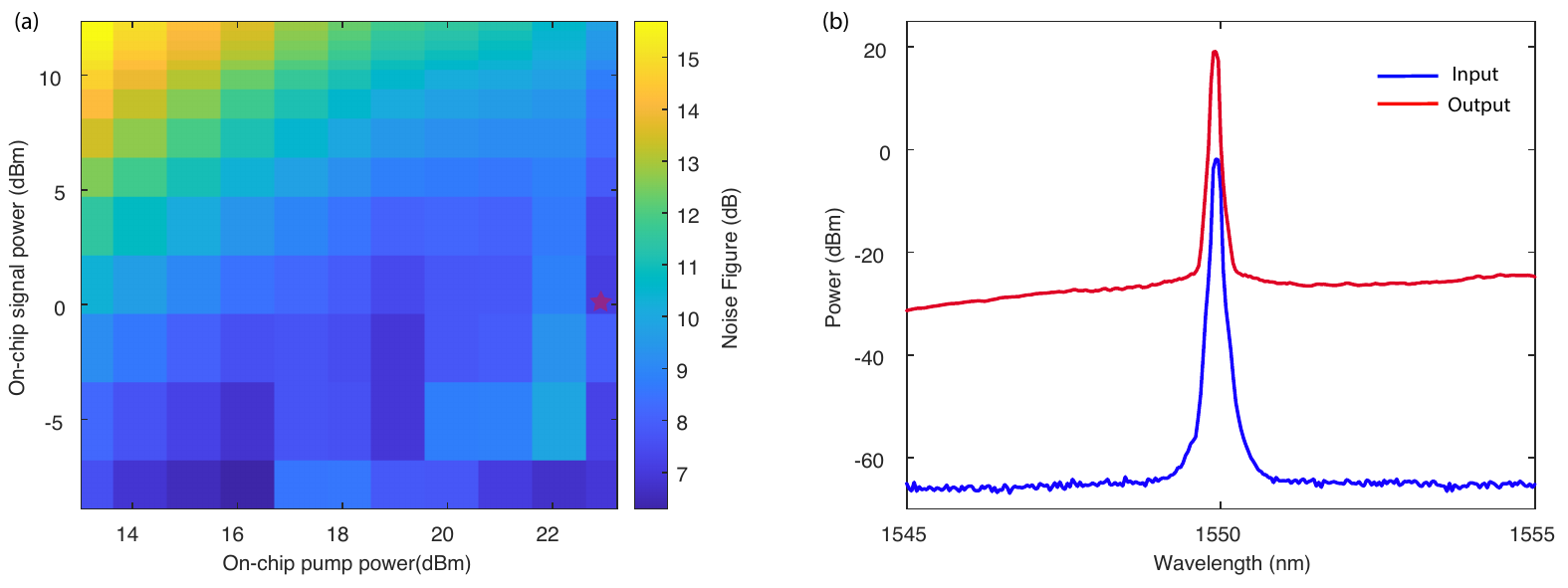}
	\caption{\textbf{Noise figure characterization}. 
	 \textbf{a},  Measured noise figure of the 21-cm-long\ce{Er\dop Si3N4} waveguide amplifier under various on-chip pump powers and off-chip input powers at 1550~nm. 
	 The star marker indicates the pump and signal conditions for the example noise figure measurement.
	 \textbf{b},  Optical spectra of the input signal and the amplified signal used for deriving the noise figure using the optical source-subtraction method\cite{Derickson1998} after power calibration.
	}
	\label{Fig:SI_NF}
\end{figure*}

The derived noise figures at various input signal powers and pump powers for a forward pumping scheme are shown in Fig.\ref{Fig:SI_NF}(a). In the high-gain regime with $>$20~dBm on-chip pump power and $<$-5 dBm off-chip signal power, the noise figure measurement is affected by the parasitic lasing effect that limits the optical gain. 
Fig. \ref{Fig:SI_NF}(b) shows an example noise figure measurement with an off-chip input signal of -1.9 dBm (marked in Fig. \ref{Fig:SI_NF}(a)).  
An off-chip optical net gain of 21~dB is obtained from the optical spectrum difference at the signal wavelength.  
The noise powers $P_\text{out}=-27$~dBm and $P_\text{SSE}=-65$~dBm over 0.2 nm optical bandwidth (the resolution bandwidth of the OSA) are interpolated from the noise floors at wavelengths 3 nm away from the signal peak, respectively. 
In order to avoid overestimate the noise figure, we use the corrected resolution bandwidth of 0.12 nm instead of the displayed  0.2 nm resolution bandwidth of the OSA. 
Using the Equation \ref{eqn_SI: NF}, a noise figure of 7.1 dB is obtained, including the contribution from the input fiber-to-chip coupling loss and relatively large spontaneous emission factor $n_\text{sp}$ upon a 1480 nm pumping (incomplete population inversion).

\section{On-chip amplification of soliton microcombs and photonic generation of microwaves.}
The soliton microcomb is generated by slow laser tuning and a single soliton is isolated using the backwards tuning technique\cite{Guo2017b}.
The generated single-soliton microcomb exhibits a characteristic sech-squared spectral envelope with an integrated power of 0.08~mW (transmitted pump is removed by fiber Bragg grating notch filters), when being excited by an on-chip pump power of around 100~mW. The measured soliton microcomb power is lower than the theoretically fitted power (0.3~mW), due to the additional optical attenuation from the chip coupling, as well as the fiber link loss between the coupled fiber and photodetector.  The fitted comb line power is given by\cite{Herr2014b}
\begin{equation}
P(\mu) = \frac{|\beta_{2}| D_{1}^{2}}{\gamma} \operatorname{sech}^{2}\left(\frac{D_{1} \mu}{\Delta \omega_{s}}\right)
\label{equation: soliton power}
\end{equation}
where $\mu$ is the comb mode number counting away from the pump line, $\beta_2$ is the group velocity dispersion coefficient, $D_1/2\pi=19.8\mathrm{GHz}$ is the cavity FSR and $\Delta\omega_s$ is the comb spectral bandwidth, yielding a maximum line power of about -29~dBm and a total microcomb power of about 0.3~mW.  Parameters used for the soliton microcomb power fitting are extracted from the reported \ce{Si3N4} Euler-bend racetrack microrsonator\cite{Ji2021}.  The group velocity dispersion $\beta_2$ is -14.1 fs$^2$/mm, corresponding to $D_{2}/2\pi=$34.9 kHz. The effective nonlinear coefficient is 0.81~W$^{-1}$m$^{-1}$. The microcomb bandwidth is given by $\delta\omega_s = 2/\pi\delta\tau_s$ where $\delta\tau_s$ is approximately 61 ps when the pump laser frequency detuning is maximized with 70 mW power in the bus waveguide. 

With the generated 19.8~GHz single-soliton state microcomb, we implement the photonic generation of microwaves using direct photodetection (Discovery Semiconductors, DSC40).
The measured single-sideband phase noise at Fourier offset frequencies of $>30$~kHz reaches a high noise plateau of -104~dBc/Hz measured by a phase noise analyzer (PNA, Rohde $\&$ Schwarz, FSW43).
This phase noise plateau is limited by photon shot noise\cite{Liang2015b} due to the low incident optical power on the photodetector (0.1~mA photocurrent).  

We use an \ce{Er\dop Si3N4} EDWA to increase the soliton power to 8.4~mW (6.9~dBm),corresponding to an off-chip net gain of 20.1~dB.
This results in 3.8~mA photocurrent, with 3.5 dB insertion loss between the chip output and the PD.  The amplified soliton is able to deliver a strong microwave signal, resulting in a reduced phase noise (in blue) for offset frequencies of $>10$~kHz, reaching -~120~dBc/Hz at 100~kHz and -126~dBc/Hz at 1~MHz, respectively.
The spectral bumps around 4 kHz offset frequencies is attributed the characteristic laser phase noise (Toptica CTL), while the spectral spikes between  10 kHz and 100 kHz stem from the laser amplitude-to-phase noise conversion observed in prior works\cite{Lucas2020,Liu2020c,Ji2021}.  
The 'step-like' phase noise at offset frequencies above 100 kHz is dominated by the phase noise analyzer noise floor (the dashed dotted line). 
We would like to note that there appear fluctuations in power of amplified comb lines from 1550 nm to 1650 nm, which stems from the wavelength-dependent enhanced amplification induced by the F-P cavity formed by chip facets.
For the case of using a commercial erbium-doped fiber amplifier,  the input soliton microcomb is amplified to a similar output power of 6.8 dBm for comparison.

\section{Transmitter OSNR budget estimation for error free short-reach transmission.}

\begin{figure*}[h]
	\centering
	\includegraphics[width=0.8\textwidth ]{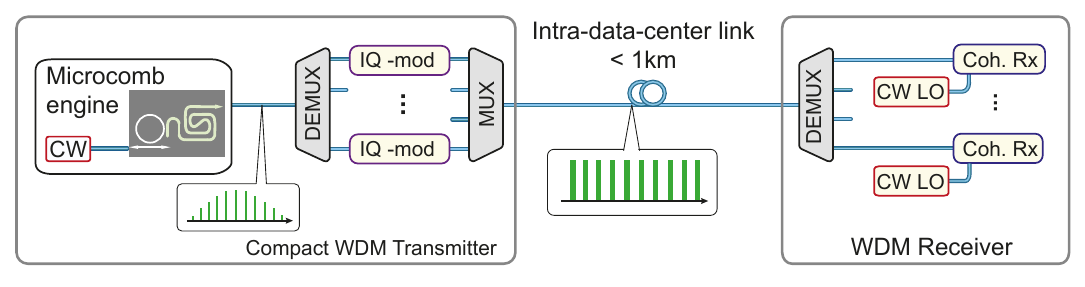}
	\caption{\textbf{Envisaged application of \ce{Er\dop Si3N4} EDWA for short-reach WDM coherent communications}.  The \ce{Er\dop Si3N4} EDWA (can be co-integrated with Kerr microrings) is used to amplify the low-power soliton microcomb with tens of mW output power for a WDM communication transmitter, exhibiting significant footprint reduction compared to the case using bench-top EDFAs. The output of the EDWA could be used for short-reach communication applications such as intra-data-center communications with $<$1 km optical fiber link.  EDWAs  can also be inserted at the output of the transmitter or the receiver front-end.
	}
	\label{Fig:SI_WDM}
\end{figure*}

Fig. \ref{Fig:SI_WDM} shows an envisaged short-reach ($<$1~km) communication application scenario, e.g. intra-data-center communications, using integrated \ce{Er\dop Si3N4} EDWAs to amplify the power of soliton microcombs. The theoretical bit error rate (BER) of 16 QAM signals can be analytically calculated from the OSNR, given by\cite{Marin-Palomo2017a}
\begin{equation}
\mathrm{BER} = \frac{3}{8}\mathrm{erfc}\sqrt{\frac{\mathrm{OSNR}}{10}},
\label{eqn:BER}
\end{equation}
where erfc is the complementary error function and OSNR is related to the measured OSNR$_\text{ref}$ by an optical spectrum analyzer (OSA), expressed by
\begin{equation}
\mathrm{OSNR} = \frac{2B_\text{ref}}{pB_{s}}\mathrm{OSNR_{ref}} ,
\end{equation}
where $B_\text{ref}$ and $B_s$ correspond to the resolution bandwidth of the OSA and the effective bandwidth of the matched filter at the receiver, $p=1(2)$ relates to the signal polarization multiplexing.
Considering a forward-error correction based on 7$\%$ redundancy, an OSNR of $>$20~dB at the receiver is required to achieve a BER less than the threshold value of $4.5\times10^{-3}$ for 16 quadrature amplitude modulation (QAM) signals at a symbol rate of 40GBd.
This sets the minimum optical signal-to-noise ratio (OSNR) requirement for the transmitter output.
For short-reach ($<$1~km) communications with negligible fiber link loss,  the output OSNRs ($>$~30 dB) of the demonstrated amplified microcomb by the \ce{Er\dop Si3N4} EDWAs can feasibly satisfy the OSNR requirement for the transmitter, allowing to tolerate around 10~dB OSNR penalty from an additional optical amplifier in the link, electronic noise, and nonlinear effects. 
\clearpage

\renewcommand{\bibpreamble}{
$\ast$These authors contributed equally to this work.\\
$\dagger$\textcolor{magenta}{yang.lau@epfl.ch}\\
$\ddag$\textcolor{magenta}{tobias.kippenberg@epfl.ch}
}
\bibliographystyle{apsrev4-2}
\bibliography{library}